\documentclass[11pt]{article}
\usepackage[margin=1in]{geometry}
\usepackage{graphicx}
\usepackage{booktabs}
\usepackage{amsmath}
\usepackage{authblk}
\usepackage[hidelinks]{hyperref}
\usepackage{caption}
\usepackage{siunitx}
\usepackage{cite}

\title{Private Noise and Public Error in Collective Information Acquisition}
\author[1,2,3]{Mohammad Salahshour\textsuperscript{*}}
\author[2]{Sumanth Bhargava}
\author[1,2,3]{Kajal Kumari}
\author[4]{Niccolo Pescetelli}
\author[5]{Yasser Roudi}
\author[6]{Bahador Bahrami\textsuperscript{\dag}}
\author[1,2,3]{Iain D. Couzin\textsuperscript{\dag}}
\affil[1]{Department of Collective Behaviour, Max Planck Institute of Animal Behavior, Konstanz, Germany}
\affil[2]{Centre for the Advanced Study of Collective Behaviour, University of Konstanz, Konstanz, Germany}
\affil[3]{Department of Biology, University of Konstanz, Konstanz, Germany}
\affil[4]{Collective Intelligence Lab, London Interdisciplinary School, London, United Kingdom}
\affil[5]{Department of Mathematics, King's College London, London, United Kingdom}
\affil[6]{Department of Psychology, Ludwig-Maximilians-Universit\"at M\"unchen, Munich, Germany}
\date{}
\begin{document}
\maketitle
\begingroup
\renewcommand{\thefootnote}{*}
\footnotetext{Correspondence: \texttt{msalahshour@ab.mpg.de}.}
\renewcommand{\thefootnote}{\dag}
\footnotetext{These authors contributed equally to this work.}
\endgroup

\begin{abstract}

Collective information acquisition requires groups to combine personal evidence with social information while remaining coupled to the external state. Communication noise can affect this process, but the role of noise remains unclear. In an online experiment, 600 participants worked in four-person human groups estimating a room temperature across 25 rounds while receiving either faithful social information, comprehension noise in which each receiver saw independently perturbed social information, or production noise in which perturbations were stored before display and could be seen by multiple receivers. The thermometer cue was objectively veridical, but its reliability was subjectively uncertain and the unitless 50--250 room-temperature range created a task-induced conflict between displayed evidence and everyday temperature expectations. Production-noise groups spent more rounds tightly clustered around a wrong value than comprehension-noise groups (\(p=0.016\), group-level permutation). Production noise more often created a wrong common signal (\(p=0.025\), Fisher's exact test) and made that signal persist across more rounds (\(p=0.004\), permutation). Dynamic update models showed that production noise was not more harmful because people followed peers more strongly, but because the same peer influence acted on more correlated production-noise perturbations. Exploratory human analyses linked the mechanism to psychological patterns while a GPT-agent experiment clarified a boundary condition: GPT agents registered uncertainty through reduced confidence without reproducing human-scale production-noise vulnerability. Overall, noise did not simply degrade collective information acquisition. Comprehension noise could sometimes improve correction relative to the faithful control, whereas production noise could turn perturbations into common evidence and stabilize consensus on error.
\end{abstract}

\section{Introduction}

Collective intelligence is often explained by error cancellation: when people bring partly independent information to a problem, averaging can improve accuracy \cite{galton1907,surowiecki2004,hong2004}. Yet this view comes with an important caveat. People do not enter social situations as blank measuring devices. They bring expectations, anchors, and prior beliefs that shape how new evidence is interpreted \cite{tversky1974,summerfield2014}. In many perceptual and judgement tasks, such priors are useful because they stabilize inference under uncertainty \cite{summerfield2014}. However, they can also bias estimates away from an available cue when that cue conflicts with familiar categories or plausible ranges \cite{tversky1974,summerfield2014}.

Social information creates a related, but distinct, vulnerability. Classic conformity experiments showed that people can move away from objectively available evidence when a group gives a conflicting answer \cite{asch1951,bond1996}. Advice-taking and wisdom-of-crowds experiments show a similar tension in quantitative judgement: seeing other people's estimates can improve accuracy, but it can also make individual errors more correlated and reduce the benefit of independent evidence \cite{yaniv2007,bahrami2010,lorenz2011,moussaid2013,jayles2017,becker2017}. Economic and computational models of herding, cascades, and networked learning explain how social signals can become self-reinforcing, sometimes causing later individuals to follow social information even when personal evidence points elsewhere \cite{bikhchandani1992,banerjee1992,andersonholt1997,weizsacker2010,degroot1974,golub2010,flache2017}. These literatures show how social signals can separate consensus from accuracy. However, they leave open a closely related question: when a prior already exists before social exchange, does communication help people revise it, or does it make the prior socially persuasive?

 The same prior-social tension becomes a collective-sensing problem when groups must estimate an external state while watching one another. Animal groups, human teams, and distributed populations often estimate an external state while also observing information generated by others facing the same environment \cite{couzin2005,dall2005,conradt2005,salahshour2019prl}. In these settings, agreement is useful only when it remains connected to the world. Groups can therefore succeed by converging on the external state, fail to coordinate, or converge on a wrong value; we call this last outcome consensus on error.

Existing social-learning experiments provide important pieces of this problem. Some ask how people combine noisy personal evidence with the observed choices of others \cite{andersonholt1997,weizsacker2010,schobel2016}; others ask whether people learn to discount social signals that are uninformative when personal evidence is valid \cite{li2019}. Less is known about situations in which people must update beliefs about an uncertain external state while balancing their own observations against the behaviour of others. The problem is visible in digital environments, where corrective information can coexist with misleading claims, perceived consensus, and collective behaviour that is poorly coupled to evidence \cite{vosoughi2018,lazer2018,bakcoleman2021}. It also arises outside digital media whenever teams or distributed populations must decide whether to trust their own local evidence or the pattern of responses around them \cite{couzin2005,dall2005,conradt2005,salahshour2019prl}. The relevant failure is therefore not always lack of truth. It can also be failure to use available evidence when prior beliefs and social signals make another interpretation feel more plausible. Social information may help correct biased anchors by exposing people to other perspectives. But if several individuals bring similar priors to the task, communication may instead turn those priors into shared confidence and, ultimately, consensus on error.

Importantly, however, communication in many real-world contexts is noisy. Messages can be misunderstood, transcribed incorrectly, summarized by platforms, distorted by intermediaries, or corrupted before being broadcast to others. In information theory, such corruption is central: communication is transmission through a channel that can alter signals \cite{shannon1948}. By contrast, many social-influence experiments treat the communication channel as if it faithfully transmits the estimates that people intend to share. Recent theory in collective sensing suggests that this simplification can miss an important effect: noise can sometimes help groups escape maladaptive agreement \cite{salahshour2019prl}, but its consequences depend on how the noise enters the communication process \cite{salahshour2019sr,Salahshour2021frontiers}.

Accounting for communication noise brings a key design question into focus: does the perturbation occur independently for each receiver under comprehension noise, or before storage and display under production noise \cite{salahshour2019sr,Salahshour2021frontiers}? In comprehension noise, a signal is produced correctly, but each receiver sees an independently perturbed version; for example, one person may misunderstand a message that others interpret correctly. In production noise, the signal is corrupted before it is stored or broadcast; for example, an erroneous value posted to a shared feed or recommended by a platform can be seen by many people as the same social fact. The same nominal amount of noise may therefore have different collective consequences: comprehension-noise perturbations are receiver-specific and can average out or weaken a wrong social attractor, whereas production-noise perturbations enter the common record and make a wrong value look socially supported.

The same design question is becoming important for human--artificial intelligence (AI) and AI--AI systems. Large language models (LLMs) are increasingly used as advisers, autonomous agents, and social simulators \cite{shanahan2023,wang2024survey,argyle2023,aher2023,park2023}. Recent work shows that LLM agents can shift their answers after receiving social or persuasive input, coordinate in multi-agent settings, and show forms of peer influence or herding \cite{griffin2023,herd2025,bringingeveryone2025}. These studies show that LLM agents are socially responsive, but much less is known about how they behave when the communication channel itself is noisy. This gap matters in environments increasingly mediated by AI agents, because such systems may not simply add information to human groups; they may filter, summarize, store, recommend, or broadcast social evidence. 

The present study brings these elements into one controlled setting. Human participants worked in four-person groups on a repeated temperature-estimation task framed as estimating the temperature of a room from a thermometer display. Participants were told that the thermometer could be noisy, and no conventional unit such as Celsius or Fahrenheit was specified. The design separates how pre-social anchors are corrected or amplified, how social communication produces informed or harmful consensus, and whether these outcomes depend on comprehension noise or production noise. A GPT-agent comparison provides a boundary-condition test for the same comprehension- and production-noise regimes. The central result is that the location of noise mattered: production noise more often turned errors into common social evidence, whereas comprehension noise was less damaging and sometimes helped loosen mistaken consensus. The mechanism was not only individual error. Shared group anchors predicted later error and bias correction, while confidence, social traits, and post-task reconstruction showed how noisy social information became psychologically consequential. GPT agents reduced confidence under noisy social information, but they did not show the same human vulnerabilities to downward anchors or to production-noise amplification. Our results suggest that noisy communication does not merely degrade collective information acquisition: depending on where it enters the channel, it can either loosen mistaken consensus or transform a biased anchor into common social evidence. They also show why this distinction matters for AI-mediated environments: GPT agents registered noisy social information as uncertainty, but did not show the same human vulnerability to downward anchors and production-noise amplification.

\section{Results}

\subsection{\texorpdfstring{Theory: comprehension noise can loosen error, production noise can stabilize it}{Theory: comprehension noise can loosen error, production noise can stabilize it}}
\label{sec:theory-results}

Collective information acquisition requires a group to remain coupled to an external state while its members also observe one another. Social information can pool partly independent observations, but it can also recycle a common error. The central failure mode is therefore not disagreement alone. It is agreement that becomes uncoupled from the state being estimated. This risk is greatest when individuals enter the interaction with expectations that make the available evidence uncertain: communication may help correct those expectations, or it may make them socially persuasive.

We formalize this problem with a minimal two-stage model. First, an individual sees evidence before social exchange and forms an initial estimate. Second, the estimate is revised after social information becomes available. This structure captures many collective-information acquisition settings: a forecaster first reads a local measurement and then sees colleagues' forecasts; a clinician first interprets a test result and then hears a team discussion; a platform user first encounters corrective information and then sees the apparent response of others.

Let \(T_t\) denote the available evidence about the external state at time \(t\), \(x_i(t)\) the estimate of individual \(i\), and \(\mu_i\) the expectation or prior value that individual \(i\) brings to the task. The pre-social estimate is
\begin{equation}
x_i(1)=\lambda_i T_1+(1-\lambda_i)\mu_i-\zeta_i,
\label{eq:initial-mixture}
\end{equation}
where \(\lambda_i\) is the evidence weight. When \(\lambda_i=1\), the estimate follows the evidence; when \(\lambda_i<1\), it is pulled toward expectation. The term \(\zeta_i\) captures additional displacement beyond this evidence--expectation mixture, such as caution or scale uncertainty. In the empirical setting below, this displacement is mostly downward, so we refer to the first-round displacement from evidence as a pre-social anchor. We use ``bias'' descriptively for signed error relative to evidence, \(x_i(t)-T_t\). We use ``pre-social anchor'' for the first-round bias before peer information appeared, interpreted as the initial displacement from evidence that social interaction could later correct or amplify. Full operational definitions are given in Methods.

Once social information is available, estimates can be pulled by two forces: personal evidence and social evidence. Let \(S_i(t)\) be the aggregate social signal available to individual \(i\). We use a linear gap-closing update because it is the simplest form that captures these forces: estimates change in proportion to their distance from current evidence and from social evidence \cite{degroot1974,yaniv2007,moussaid2013,jayles2017}.
\begin{equation}
x_i(t+1)-x_i(t)=\omega_i(t)\,[T_t-x_i(t)]+\sigma_i(t)\,[S_i(t)-x_i(t)]+\epsilon_i(t),
\label{eq:dynamic-update}
\end{equation}
Here \(\omega_i(t)\) is the weight on personal evidence and \(\sigma_i(t)\) is the weight on social information. The term \(T_t-x_i(t)\) is evidence pull; \(S_i(t)-x_i(t)\) is peer pull. Empirically, the coefficient on peer pull estimates \(\sigma_i(t)\). When an empirical coefficient estimates a named theory parameter, we report it with a hat, for example \(\widehat{\sigma}_C\); \(\beta\) is reserved for regression coefficients that do not correspond to a named model parameter. Throughout the paper, \emph{comprehension noise} denotes independently perturbed social information at reception, whereas \emph{production noise} denotes perturbation before storage or display. These are the only condition labels used below.

We treat \(\sigma_i(t)\) not as a fixed conformity parameter, but as a testable social-information weight. In a general collective-information acquisition problem, the pull of social information should depend on cues to how reliable and useful that social information is. A noisy communication channel can make the social signal less trustworthy; if the error is introduced before a message is shared, the same perturbed value can also be repeated as common evidence, as in social-learning and cascade settings \cite{bikhchandani1992,andersonholt1997}. Confidence can matter because an individual who trusts their current estimate has less reason to close the gap toward the social signal \cite{yaniv2007,bahrami2010}. Peer dispersion can matter because a scattered social record is less diagnostic than a coherent one \cite{koriat2012}. Anchors can matter because an initial displacement from evidence may change whether a person or group treats the social record as corrective or confirmatory \cite{tversky1974,summerfield2014}. We collect these candidate moderators in a deliberately testable form as follows:
\begin{equation}
\begin{aligned}
&\\
\sigma_i(t)&=\sigma_0+\sigma_N I_{\mathrm{noise}}+\sigma_P I_{\mathrm{prod}}+\sigma_C C_i(t)+\sigma_D D_i(t)+\sigma_G A_g+\sigma_a A_i,
\end{aligned}
\label{eq:social-weight}
\end{equation}

where \(C_i(t)\) is confidence, \(D_i(t)\) is peer dispersion, \(A_g\) is the shared group anchor, and \(A_i\) is the individual's own pre-social anchor magnitude. Here ``noise'' means perturbation in communicated social information, not uncertainty in the evidence \(T_t\). Thus \(I_{\mathrm{noise}}\) marks a noisy social channel, and \(I_{\mathrm{prod}}\) tests the additional production-noise component beyond comprehension noise. Each term in Eq.~\ref{eq:social-weight} can be tested as an interaction with peer pull, \(S_i(t)-x_i(t)\). If \(\sigma_C<0\), confidence reduces the social-information weight; if \(\sigma_D<0\), peer dispersion reduces the social-information weight; if \(\sigma_G>0\), shared anchors increase the social-information weight; and if \(\sigma_a>0\), personal anchors do the same at the individual level. These terms are therefore candidate mechanisms to be tested, not assumptions that the corresponding effects must exist.

We include confidence in the model because people often use confidence as a cue to whether they should rely on their own estimate or give more weight to others \cite{yaniv2007,bahrami2010,koriat2012}. Confidence should therefore vary with cues to reliability: it may be lower when the estimate is far from the evidence, when peers disagree, when the social channel is noisy, or when the initial anchor makes the evidence hard to interpret. It may also be higher when the group record appears coherent, even if that coherence is biased. Equation~\ref{eq:confidence-state} formalizes this reliability-check layer:
\begin{equation}
C_i(t)=\kappa_0+\kappa_N I_{\mathrm{noise}}+\kappa_E |x_i(t)-T_t|+\kappa_D D_i(t)+\kappa_a A_i+\kappa_g A_g+\upsilon_i(t),
\label{eq:confidence-state}
\end{equation}
Here \(\kappa_N\) tests the association between noisy social channels and confidence, \(\kappa_E\) the association between current error and confidence, \(\kappa_D\) the association between peer dispersion and confidence, and \(\kappa_a\) and \(\kappa_g\) the associations of personal and shared anchors with confidence. The residual term \(\upsilon_i(t)\) captures the remaining variation in confidence. If confidence is a reliability signal, noisy channels, larger error, greater peer dispersion, and stronger personal anchors should tend to lower it. The shared-anchor term is left open because a shared bias could either reduce confidence by conflicting with the evidence or increase confidence by making the social record appear coherent. Eq.~\ref{eq:confidence-state} tests whether confidence varies with these cues; by itself it neither proves nor assumes that changing confidence would causally change later updating.

The channel then affects the update in a different way: it changes the structure of the social signal \(S_i(t)\), not only, possibly, the size of \(\sigma_i(t)\). Under comprehension noise, each receiver gets an independently perturbed social signal. Under production noise, the perturbation is applied before storage or display, so the perturbed estimate enters the common record,
\begin{align}
S_i^{C}(t)&=\frac{1}{N-1}\sum_{j\ne i}\big[x_j(t)+\eta_{ij}^{C}(t)\big],
\label{eq:comprehension-signal}\\
S_i^{P}(t)&=\frac{1}{N-1}\sum_{j\ne i}\big[x_j(t)+\eta_j^{P}(t)\big].
\label{eq:production-signal}
\end{align}
This is our formal noise-location prediction. It follows from the update equation (Eq.~\ref{eq:dynamic-update}) and the structure of \(S_i(t)\): even if the social-information weight \(\sigma_i(t)\) were held fixed, comprehension and production noise would feed different social signals into the same update rule. In comprehension noise, the perturbation is receiver-specific; in production noise, it is attached to the peer estimate before that estimate enters the shared record. The prediction therefore does not require the confidence equation or the moderator terms in Eq.~\ref{eq:social-weight}: in a minimal version of the model, those terms can be left unspecified and the formal noise-location result follows from Eq.~\ref{eq:dynamic-update} and Eqs.~\ref{eq:comprehension-signal}--\ref{eq:production-signal}. As derived in Methods, this difference gives production noise \(N-1\) times more group-level channel variance than comprehension noise and makes receivers' noisy social signals correlated. This covariance structure is what allows production noise to turn one perturbation into shared evidence, whereas comprehension-noise perturbations are more likely to remain idiosyncratic and average out.

This theory yields the empirical predictions summarized in Table~\ref{tab:theory-predictions}. The first three are direct update predictions: an anchor-compression prediction, an evidence-pull prediction, and a peer-pull prediction. They ask whether initial estimates are compressed toward expectation, whether later estimates move toward evidence, \(T_t-x_i(t)\), and whether they move toward the social signal, \(S_i(t)-x_i(t)\). Predictions 4 and 7--9 are social-weight predictions: they ask whether channel noise, confidence, peer dispersion, and anchors change the social-information weight, \(\sigma_i(t)\). Predictions 5--6 are noise-location predictions: they ask whether production noise creates a more correlated channel-noise component and whether that component moves the group most clearly when it enters the common record. Prediction 10 is a correctability prediction: comprehension noise can help when biased anchors are still recoverable. Prediction 11 is a confidence-state prediction: confidence should fall when reliability cues worsen, while shared anchors are treated as an open context term.
Hereafter, SI denotes the Supplementary Information; SI Table and SI Fig. denote supplementary tables and figures.

\begin{table}[!t]
\centering
\caption{Theory predictions and empirical tests. The Status column distinguishes direct consequences of the update equations, consequences of the channel covariance structure, and moderator or confidence-state hypotheses. Coefficients are cluster-robust OLS estimates unless otherwise noted.}
\label{tab:theory-predictions}
\tiny
\setlength{\tabcolsep}{2pt}
\renewcommand{\arraystretch}{0.88}
\begin{tabular}{p{0.19\textwidth}p{0.09\textwidth}p{0.26\textwidth}p{0.36\textwidth}}
\toprule
Prediction & Status & Theoretical ground & Empirical evidence \\
\midrule
1. Initial estimates should be compressed toward expectation when evidence conflicts with prior belief. & Direct & Eq.~\ref{eq:initial-mixture}: if \(\lambda_i<1\) and \(\mu_i<T_1\), high evidence is pulled downward toward expectation. & Round-1 slopes were below 1 in all regimes: control \(0.807\), comprehension \(0.762\), production \(0.765\); all slope-below-one tests \(p\leq0.0089\) (Fig.~\ref{fig:bias-individual}(a)). \\
2. Estimates should show evidence pull, \(T_t-x_i(t)\), when personal evidence remains usable. & Direct & Positive personal-evidence weight, \(\omega_i(t)>0\), in Eq.~\ref{eq:dynamic-update}. & Evidence-pull coefficients, \(T_t-x_i(t)\), estimating the personal-evidence weight were small after previous estimates and peer pull, \(S_i(t)-x_i(t)\), were included: control \(\widehat{\omega}=0.038\), \(p=0.082\); comprehension \(\widehat{\omega}=-0.002\), \(p=0.883\); production \(\widehat{\omega}=0.045\), \(p=0.321\). Thus, later direct recoupling to evidence was weaker than peer pull (SI Table~S67). \\
3. Estimates should move toward social information, \(S_i(t)\). & Direct & Positive social-information weight, \(\sigma_i(t)>0\), on peer pull, \(S_i(t)-x_i(t)\), in Eq.~\ref{eq:dynamic-update}. & The social-information weight, \(\sigma_i(t)\), estimated as the coefficient on peer pull, \(S_i(t)-x_i(t)\), was positive in all regimes: control \(\widehat{\sigma}=0.266\), \(p=4.1\times10^{-5}\); comprehension \(\widehat{\sigma}=0.507\), \(p=3.6\times10^{-15}\); production \(\widehat{\sigma}=0.502\), \(p=9.2\times10^{-15}\) (Fig.~\ref{fig:dynamic-update}(a)). \\
4. Noisy channels can increase the social-information weight, \(\sigma_i(t)\), without a production-specific increase. & Moderator & Eq.~\ref{eq:social-weight}: \(\sigma_N\) allows noisy channels to change the social-information weight, \(\sigma_i(t)\), while \(\sigma_P\) tests production beyond comprehension. & In a condition-only interaction model (with comprehension-noise and production-noise indicators as the only peer-pull moderators), the comprehension-control increment estimated the noisy-channel term \(\widehat{\sigma}_N=0.240\) (\(p=0.0082\)). The production-comprehension increment estimated the production-specific term \(\widehat{\sigma}_P=-0.004\) (\(p=0.963\)), so the total production-control increment was \(\widehat{\sigma}_N+\widehat{\sigma}_P=0.236\) (\(p=0.0097\)). Thus, both noisy conditions increased \(\sigma_i(t)\), but production noise did not add a larger social-information weight beyond comprehension noise. In the fuller moderator model, the residual noisy-channel term was not reliable after confidence, \(C_i(t)\), peer dispersion, \(D_i(t)\), personal anchors, \(A_i\), and shared group anchors, \(A_g\), were entered together, suggesting that these state and context variables accounted for part of the condition-only noisy-channel increment (SI Section 12.1). \\
5. Production noise should create more correlated channel-noise components. & Channel & Eqs.~\ref{eq:variance-ratio}--\ref{eq:receiver-correlation}; comprehension-noise perturbations average out, whereas production-noise perturbations are reused across receivers. & Observed group-mean channel variance was \(0.841\) in comprehension noise and \(2.228\) in production noise; receiver-to-receiver correlations were \(0.004\) and \(0.640\), respectively. Because this pattern follows from the implemented channel structure, the formal covariance check is reported in SI Section 12. \\
6. Group-level channel-noise components should move the group most clearly under production noise. & Channel & Production-noise perturbations survive averaging and enter the group-level social signal. & Group-level channel-noise components predicted next group-mean movement in production noise (\(\beta=0.389\), \(p=0.016\)) but not comprehension noise (\(\beta=-0.009\), \(p=0.961\)); interaction \(p=0.083\) (Fig.~\ref{fig:dynamic-update}(d)). \\
7. Confidence should reduce the social-information weight, \(\sigma_i(t)\). & Moderator & In Eq.~\ref{eq:social-weight}, \(\sigma_C<0\) means confidence reduces the social-information weight. & Higher previous confidence reduced the social-information weight, \(\sigma_i(t)\), in the pooled full model (\(\widehat{\sigma}_C=-0.130\), \(p=3.5\times10^{-7}\)) and within all regimes (all \(p\leq0.048\); SI Tables~S74 and~S70; SI Section 12.1). \\
8. Peer dispersion, \(D_i(t)\), should reduce the social-information weight, \(\sigma_i(t)\), if disagreement marks low social reliability. & Moderator & In Eq.~\ref{eq:social-weight}, \(\sigma_D<0\) means peer dispersion reduces \(\sigma_i(t)\). & The peer-pull by dispersion interaction, \([S_i(t)-x_i(t)]\times D_i(t)\), estimated \(\widehat{\sigma}_D=-0.063\) in the pooled full model (\(p=0.044\)) and was especially negative in production noise (\(\widehat{\sigma}_D=-0.152\), \(p=1.8\times10^{-25}\)); it was not reliable in control and was weaker, though reliable, in comprehension noise. We therefore treat this as a qualified moderator result (SI Section 12.1). \\
9. Shared and personal anchors may differ in how they change the social-information weight, \(\sigma_i(t)\). & Moderator & In Eq.~\ref{eq:social-weight}, \(\sigma_G\) and \(\sigma_a\) test whether the shared group anchor or the personal anchor changes the social-information weight. & After both anchor terms were entered, shared group anchor still increased the social-information weight, \(\sigma_i(t)\), (\(\widehat{\sigma}_G=0.051\), \(p=0.00043\)), whereas personal anchor did not (\(\widehat{\sigma}_a=0.000009\), \(p=0.9996\)). Shared anchors also predicted later error more strongly than personal anchors (Fig.~\ref{fig:bias-individual}(d); SI Section 12.1). \\
10. Comprehension noise can help when anchors are correctable. & Channel & Eqs.~\ref{eq:bias-update}--\ref{eq:faithful-bias-record}; faithful transmission carries current group bias, whereas comprehension-noise perturbations weaken the coherence of biased peer pull, \(S_i(t)-x_i(t)\), while evidence pull, \(T_t-x_i(t)\), remains. & At a representative \(5^\circ\) anchor, model-implied correction was near zero in comprehension noise (\(-0.02^\circ\)) but negative in control (\(-4.84^\circ\), \(p=0.017\) vs comprehension) and production noise (\(-5.05^\circ\), \(p=0.0015\) vs comprehension; Fig.~\ref{fig:bias-individual}(h)). \\
11. Confidence should fall when reliability cues worsen. & Confidence state & Eq.~\ref{eq:confidence-state}: noisy channels, larger error, peer dispersion, \(D_i(t)\), and stronger personal anchors can reduce confidence, while the shared group anchor is estimated as a context term. & In the pooled confidence equation, noisy channels reduced confidence (\(\widehat{\kappa}_N=-0.552\), \(p=0.015\)), peer dispersion, \(D_i(t)\), reduced confidence (\(\widehat{\kappa}_D=-0.278\), \(p=0.043\)), and personal anchor magnitude reduced confidence (\(\widehat{\kappa}_a=-0.223\), \(p=0.049\)); current error was negative but marginal overall. Shared group anchor was positively associated with confidence, so this part of the confidence equation is descriptive rather than a directional reliability prediction (SI Section 12.1). \\
\bottomrule
\end{tabular}
\end{table}

\subsection{\texorpdfstring{Experiment and analysis overview}{Experiment and analysis overview}}

\begin{figure}[p]
 \centering
 \includegraphics[width=\textwidth]{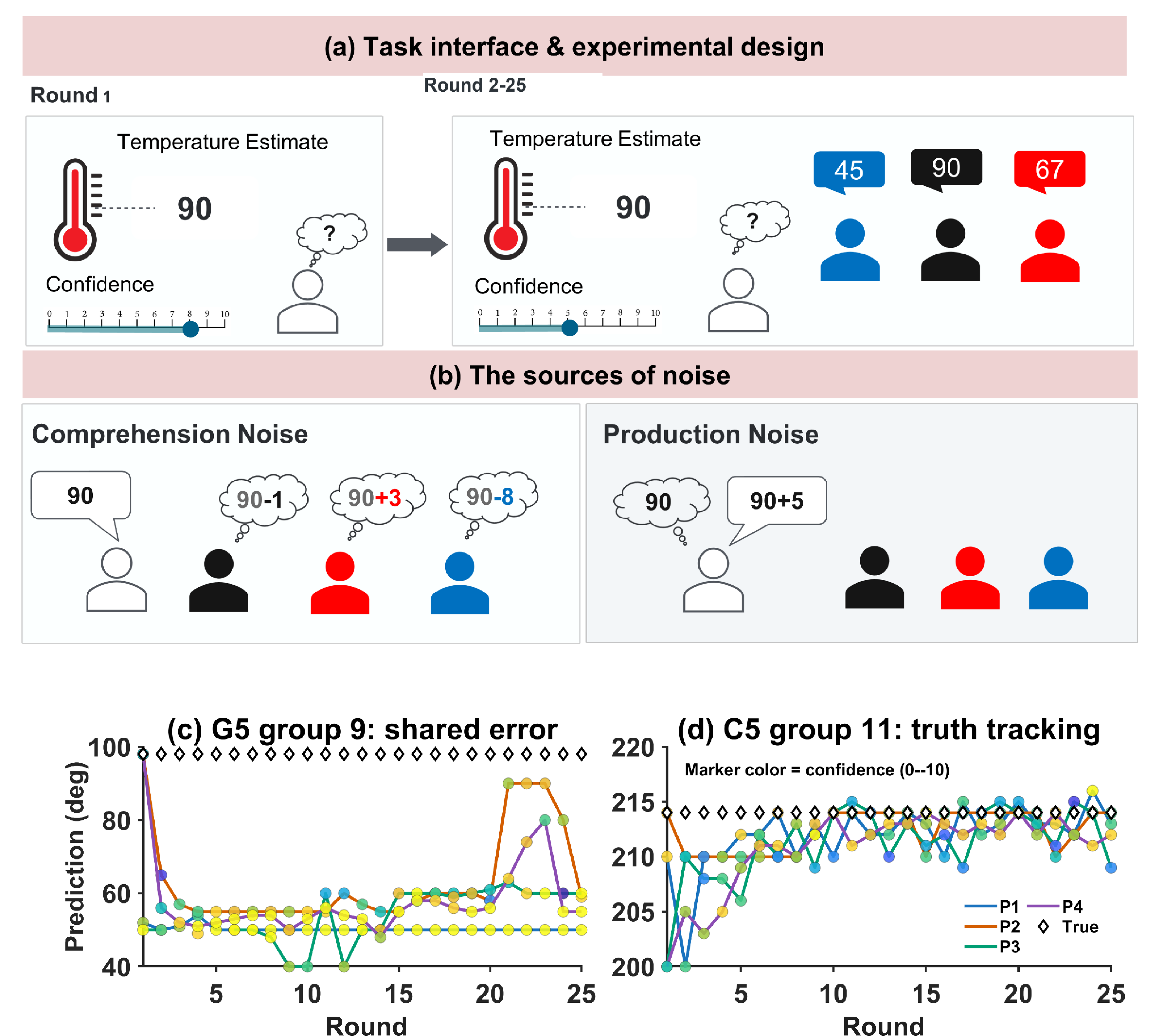}
 \caption{Experimental design, communication manipulation, and example trajectories. Panel (a) illustrates the repeated temperature-estimation task: participants reported a temperature estimate and confidence, and from round 2 onward saw social information from other group members' previous estimates. Panel (b) illustrates the two noise locations. In the control condition, previous estimates were displayed without added social distortion. In comprehension noise, each receiver saw an independently perturbed version of the same previous estimate. In production noise, the estimate was perturbed before storage and could therefore become shared social evidence for multiple receivers. Panels (c),(d) show representative group trajectories. Lines connect each participant's estimates across rounds; line color identifies participants, marker color shows confidence, and black diamond markers show the true temperature. Panel (c) shows production-noise group 9, which converged toward a shared but inaccurate estimate; panel (d) shows comprehension-noise group 11, which tracked the true temperature. All group trajectories are presented in the SI (Section 11 and SI Figs.~S36--S44).}
 \label{fig:design}
 \label{fig:example-trajectories}
\end{figure}

The human experiment tested these predictions in 600 online participants assigned to four-person groups (Fig.~\ref{fig:design}(a),(b)). Participants saw a thermometer cue before each estimate and were told that the thermometer could be noisy; objectively, however, the thermometer display equaled the true room temperature in all rounds. The room temperature was drawn uniformly at random from the interval \([50,250]\) for each group and held fixed across rounds. The task therefore placed a veridical but subjectively uncertain evidence source in competition with prior expectations and social information.

From round 2 onward, participants also saw peers' previous estimates, providing the social signal. The control condition transmitted those estimates faithfully. Comprehension noise followed Eq.~\ref{eq:comprehension-signal}: each receiver saw independently perturbed social information. Production noise followed Eq.~\ref{eq:production-signal}: an estimate was perturbed before storage and broadcast, allowing the corrupted value to become part of the shared social record. In figures and tables, C5 denotes the comprehension-noise condition and G5 denotes the production-noise condition; the \(5\) indicates that social-channel perturbations were sampled uniformly from \([-5,5]\) task units. This design asks whether groups remain coupled to available evidence when the social record is faithful, affected by comprehension noise, or affected by production noise.

Within the notation of Section~\ref{sec:theory-results}, the external state was a room temperature, the evidence \(T_t\) was a thermometer display, and \(x_i(t)\) was participant \(i\)'s estimate. The expectation term \(\mu_i\) corresponds to each participant's prior sense of plausible room temperatures, and \(S_i(t)\) corresponds to the social signal. Because the sampled room temperature could affect task difficulty and anchor strength, we checked whether the randomly assigned groups differed in their true temperatures across conditions. They did not differ reliably (group means: control 157.9, comprehension noise 137.4, production noise 149.4; Kruskal--Wallis \(p=0.172\)). The primary consensus model nevertheless adjusts for room temperature and shared anchor structure, so the main group-level comparison is not based only on the unadjusted randomization check.

We also ran a matched GPT-agent experiment as a boundary-condition analysis. The GPT agents received the same task framing and social-channel conditions, but they are used here to ask how artificial agents respond to the same information structure, not to claim that they are direct models of human collective sensing. The exact prompts, run counts, sampling temperatures, persistent histories, parsing rules, and post-task questions are specified in Methods and the SI. Detailed human and GPT procedures, variable definitions, incentives, post-task questionnaires, and statistical models are provided in Methods and in the SI.

We organize the results to follow this evidential chain. Section~\ref{sec:individual-effects} first reports individual-level condition effects on error, confidence, disagreement, and social-signal inconsistency. Section~\ref{sec:initial-bias} then examines pre-social anchors and bias correction. Section~\ref{sec:harmful-consensus} tests participant-level harmful consensus and group-level misinformed consensus, with group-level duration as the primary behavioral outcome. Section~\ref{sec:dynamic-updates} tests the dynamic mechanism by estimating evidence pull, peer pull, and channel-noise components. The remaining results treat post-task reconstruction, confidence and social traits, demographic heterogeneity, and GPT agents as exploratory or boundary-condition analyses, before the mechanistic synthesis returns to the main interpretation.

To keep the evidential hierarchy explicit, we treat group-level misinformed-consensus duration based on submitted estimates as the primary behavioral outcome. Misinformed consensus in the noised shared social record is the mechanistic channel outcome, because it describes what production noise made available as common evidence. Participant-level prediction error, confidence, and social-signal inconsistency are convergent individual-level outcomes, whereas post-task reconstruction, traits, demographics, and GPT agents are exploratory or boundary-condition extensions.

\begin{table}[t]
\centering
\caption{Evidential hierarchy for the main analyses.}
\label{tab:evidence-hierarchy}
\footnotesize\begin{tabular}{p{0.30\textwidth}p{0.28\textwidth}p{0.17\textwidth}p{0.15\textwidth}}
\toprule
Claim & Outcome & Unit & Status \\
\midrule
Production noise increases consensus on error & Misinformed-consensus duration, submitted estimates & Group & Primary \\
Production noise corrupts common evidence & Misinformed-consensus entry and duration, shared social record & Group/channel & Mechanistic \\
Participants follow the social signal, \(S_i(t)\) & Dynamic update weight & Participant-round & Mechanistic \\
Anchors seed later error & Round-1 personal, peer, and group anchors & Participant/group & Mechanistic \\
Reconstruction, traits, demographics, GPT agents & Post-task, survey, demographic, and agent outcomes & Mixed & Exploratory \\
\bottomrule
\end{tabular}
\end{table}

To make the task dynamics concrete, Fig.~\ref{fig:design}(c),(d) shows two example groups. Each panel shows the predictions of the four participants across rounds; line color identifies participants, marker color shows confidence, and black diamond markers show the true temperature. In both examples, participants within a group converged toward similar predictions; however, one group approached the true temperature, whereas the other drifted away from it. They preview the central empirical problem: social convergence was relatively common, but it was not always truthful, even though participants repeatedly received an objectively accurate, yet subjectively uncertain evidence cue. All group trajectories are presented in the SI (Section 11 and SI Figs.~S36--S44).

\subsection{\texorpdfstring{Noise effects were channel-specific rather than uniformly harmful}{Noise effects were channel-specific rather than uniformly harmful}}
\label{sec:individual-effects}

\begin{figure}[p]
 \centering
 \includegraphics[width=\textwidth]{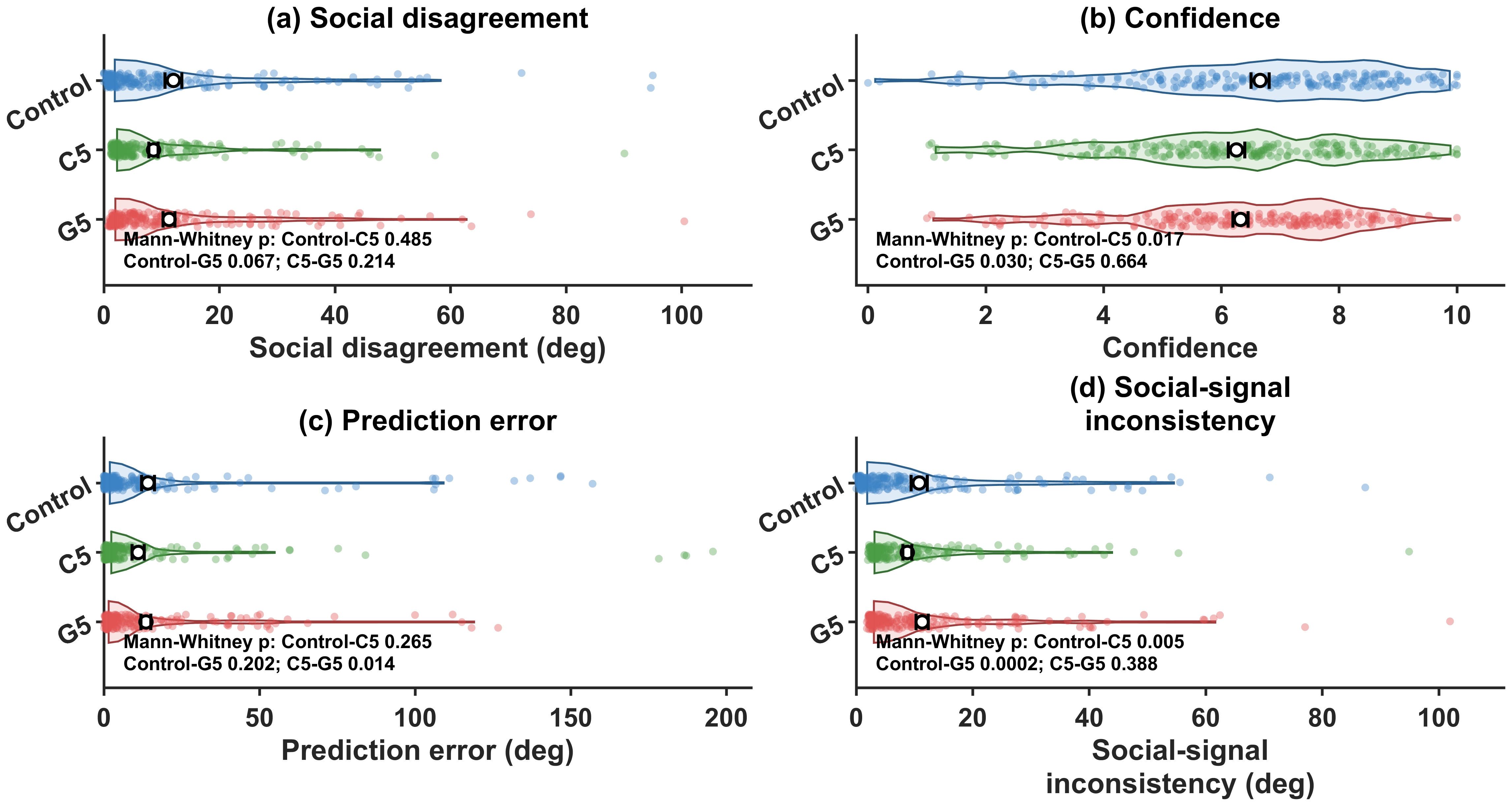}
 \caption{Human condition effects. Panels (a)--(d) show participant-level averages across rounds for social disagreement, confidence, prediction error, and social-signal inconsistency. Social-signal inconsistency is the mean absolute distance between the submitted estimate, \(x_i(t)\), and the social signal, \(S_i(t)\). Transparent violin summaries and participant-level points are overlaid on the same condition row, with the points shown more prominently. White markers show means with standard errors. P-value lines report two-sided pairwise Mann--Whitney tests on participant-level averages. Group-mean and cluster-robust robustness checks are reported in SI Table~S2.}
 \label{fig:human-effects}
\end{figure}

We begin with the individual-level behavioral summaries. Noisy communication did not simply increase disagreement: social disagreement was lowest in comprehension noise, but pairwise tests did not support a monotonic noise effect (Fig.~\ref{fig:human-effects}(a); SI Section 3; SI Table~S2). Confidence showed a clearer response to the manipulation. Both noisy conditions reduced confidence relative to control, indicating that participants registered the uncertainty introduced by the social channel (Fig.~\ref{fig:human-effects}(b); SI Table~S2). This confidence reduction reappears in the dynamic confidence-state model in Section~\ref{sec:dynamic-updates} (the noisy-channel term \(\kappa_N\) in Eq.~\ref{eq:confidence-state} predicted lower \(C_i(t)\)).

Prediction error showed the main comprehension-production asymmetry. Participants in production noise were less accurate than participants in comprehension noise (\(p=0.014\), Mann--Whitney test; Fig.~\ref{fig:human-effects}(c); SI Table~S2), whereas neither noisy condition differed reliably from control. Social-signal inconsistency, defined as the distance from the social signal, \(S_i(t)\), did not differ reliably between the two noisy treatments (Fig.~\ref{fig:human-effects}(d)). Thus, production noise was not more harmful because participants were generally less responsive to peer pull, \(S_i(t)-x_i(t)\). Rather, the same social channel became more dangerous when corrupted values entered the shared record. This individual-level pattern foreshadows the dynamic result in Section~\ref{sec:dynamic-updates}: as we will see, the social-information weight, \(\sigma_i(t)\), was similar in comprehension and production noise, but production-noise perturbations in \(S_i(t)\) were more able to move the group state.

Because participants were nested in four-person groups, we treat these participant-level summaries as convergent evidence rather than the primary collective test. Group-mean, cluster-robust, and permutation analyses preserved the direction of the prediction-error contrast but were more conservative (SI Section 4; SI Tables~S2--S5). The collective claim is therefore evaluated primarily through the group-level consensus, anchor, and time-course analyses below, where the production-noise signature was clearer (SI Tables~S3--S15).

\subsection{\texorpdfstring{Initial bias shaped later individual and group behavior}{Initial bias shaped later individual and group behavior}}
\label{sec:initial-bias}

Later errors were not created by social influence alone. Participants entered the task with downward anchors before seeing any peer estimates. This is the empirical counterpart of the first stage of the theory: the initial estimate in Eq.~\ref{eq:initial-mixture} can already be displaced from evidence before any peer signal is available. Furthermore, anchors persisted across individuals, groups, and countries. Notably, the group bias predicted later error more strongly than personal anchors, while personal anchors were more closely tied to confidence. These initial anchors also changed differently over time: comprehension noise corrected weak anchors most consistently, whereas production noise was most risky when weak anchors could become shared social evidence.

\subsubsection{\texorpdfstring{Participants showed pre-social downward anchors}{Participants showed pre-social downward anchors}}

\begin{figure}[p]
 \centering
 \includegraphics[width=\textwidth]{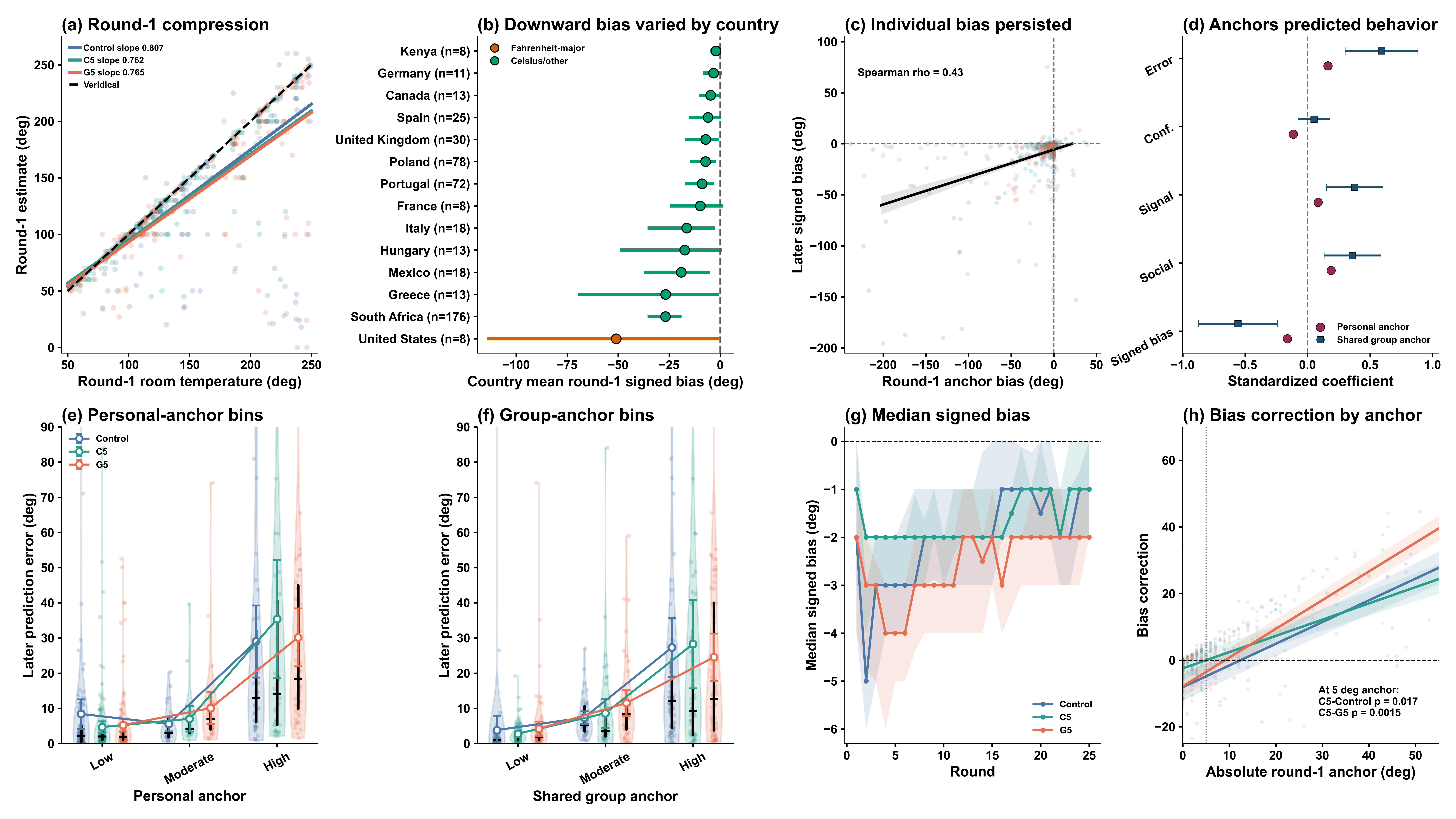}
 \caption{Anchor bias, persistence, and correction. Panel (a) shows round-1 prediction against round-1 actual room temperature with condition-specific fits; the dashed line is veridical responding. Panel (b) shows country-level mean round-1 signed anchor bias for countries with at least eight matched participants; intervals are Bayesian-bootstrap 95\% credible intervals, and color marks the broad temperature-scale context. Panel (c) shows the participant-level relation between round-1 anchor bias and later signed bias. Panel (d) compares personal and shared group anchors in condition-adjusted cluster-robust models; Error denotes prediction error, Conf. confidence, Signal social-signal inconsistency, Social social disagreement, and Signed bias later signed prediction bias. Panels (e,f) show how later prediction error increased across personal-anchor and group-anchor bins; low, moderate, and high anchor bins are \(\leq5^\circ\), \(>5^\circ\)--\(12^\circ\), and \(>12^\circ\). Panel (g) shows median signed bias over rounds; the median is used because raw round-level means were sensitive to a few extreme typed responses documented in SI Fig.~S27 and SI Tables~S44--S45. Panel (h) shows bias correction relative to the round-1 anchor and annotates the key HC3 OLS contrasts at \(5^\circ\) anchor magnitude. Country-level persistence is reported in SI Section 7.1. Model and test details are reported in SI Sections 7.1--7.8 and SI Tables~S46, S37--S39, and~S44--S45.}
 \label{fig:bias-individual}
 \label{fig:bias-individual-descriptive}
\end{figure}

An interesting feature of the participants' behavior was that their errors were directional, not merely large or small: participants showed a systematic signed prediction bias (\(x_i-T_t\)) toward lower temperatures. Across rounds, the participant-level signed prediction bias averaged \(-11.00^\circ\) across all participants. The first round is especially informative because it precedes any social exposure. Pre-social anchor bias was already negative in control (\(-16.99^\circ\), sign-rank \(p=5.8\times10^{-15}\)), comprehension noise (\(-13.87^\circ\), \(p=1.9\times10^{-16}\)), and production noise (\(-18.41^\circ\), \(p=2.3\times10^{-16}\)), and these pre-social condition differences were not significant (all pairwise Mann--Whitney tests \(p\geq0.091\)). Round-1 predictions increased with actual temperature, but sublinearly in every condition (Fig.~\ref{fig:bias-individual}(a)): the slope was 0.807 in control (95\% confidence interval (CI) [0.662, 0.952], \(p=0.0089\) for slope \(<1\)), 0.762 in comprehension noise (95\% CI [0.634, 0.890], \(p=2.6\times10^{-4}\)), and 0.765 in production noise (95\% CI [0.651, 0.878], \(p=5.0\times10^{-5}\)). The between-condition slope differences were not significant (control vs comprehension \(p=0.667\), production vs comprehension \(p=0.977\), control vs production \(p=0.608\); model details are provided in SI Section 7.1). Thus, participants did not simply add noise around the truth. Before any communication manipulation took effect, they already compressed high temperatures downward, consistent with an anchor in which the higher temperatures shown in the task could seem implausibly hot for a room. This interpretation is deliberately broader than a pure-prior account. Because room temperatures were drawn uniformly at random between 50 and 250, the upper part of the range could conflict with everyday expectations about plausible rooms; because participants were also told that the thermometer could be noisy and the scale had no conventional unit label, the pre-social anchor is best understood as a task-induced conflict among displayed evidence, scale ambiguity, cue reliability, and real-world expectations. The experiment did not directly ask participants how they interpreted the scale or thermometer reliability, so this mechanism is inferred from behavior rather than directly measured.

The same downward bias was visible across major countries, but its magnitude varied substantially, suggesting cultural differences in how participants mapped the task onto plausible room-temperature expectations (Fig.~\ref{fig:bias-individual}(b)). Country mean round-1 pre-social anchor bias was directionally associated with country mean later signed bias (SI Section 7.1).

We also asked whether this pattern reduced to a simple Celsius-versus-Fahrenheit contrast within countries. It did not: the small Fahrenheit-major subgroup was, if anything, more negatively biased than the Celsius/other subgroup (SI Fig.~S19(b)). A more plausible reading is that participants imported broader real-world expectations about plausible room temperatures, and those expectations differed across countries and individuals.

\subsubsection{\texorpdfstring{Participants' initial anchors predicted later bias and error}{Participants' initial anchors predicted later bias and error}}

We then asked whether later performance tracked each participant's own initial anchor or the shared anchor already present in the group. At the individual level, the pre-social anchor clearly predicted later behavior: participants with more negative round-1 anchors also showed more negative signed estimates later in the task (Spearman \(\rho=0.43\), \(p=3.6\times10^{-27}\); Fig.~\ref{fig:bias-individual}(c)). This supports the idea that later underestimation was not generated anew by social interaction alone; it was seeded by prior expectations already present on the first round.

\subsubsection{\texorpdfstring{Shared group anchors predicted behavior, whereas personal anchors predicted confidence}{Shared group anchors predicted behavior, whereas personal anchors predicted confidence}}
\label{sec:shared-personal-anchors}

At the group level, we defined the shared group anchor as the absolute value of the group's mean signed round-1 bias. Group anchors were smaller than the average absolute individual anchor because averaging partly cancelled opposing individual biases (SI Fig.~S23; SI Table~S37).

The shared group anchor was a stronger predictor of later choices than individuals' own anchor. In pooled cluster-robust models that entered personal anchor, group anchor, and condition together, group anchor predicted larger later prediction error more strongly than personal anchor (\(\beta=0.592\) vs \(0.162\), coefficient-difference \(p=0.0083\); Fig.~\ref{fig:bias-individual}(d); SI Section 7.4 and SI Table~S38). This condition-adjusted comparison shows that the shared group anchor carried information beyond each participant's own initial bias. The same pooled model showed that group anchor was also the stronger predictor of social-signal inconsistency (\(\beta=0.376\) vs \(0.084\), coefficient-difference \(p=0.0105\)) and later signed bias (\(\beta=-0.558\) vs \(-0.162\), coefficient-difference \(p=0.0225\)), whereas personal anchor was the only reliable predictor of lower confidence (\(\beta=-0.115\), \(p=0.044\); group anchor \(\beta=0.051\), \(p=0.431\)).

This dissociation suggests that shared group anchors and personal anchors played different roles: the shared anchor shaped later collective error, whereas confidence remained tied to each participant's own initial anchor. Thus, the bias that survived aggregation was smaller than the average individual bias, but it was more consequential for later behavior. The dynamic analyses later give the same distinction a process interpretation: the shared group anchor, \(A_g\), increased the social-information weight, \(\sigma_i(t)\), whereas the personal anchor, \(A_i\), did not (Section~\ref{sec:dynamic-updates}).

\subsubsection{\texorpdfstring{Shared group anchors mattered most under comprehension noise}{Shared group anchors mattered most under comprehension noise}}

We then asked whether this group-over-personal pattern held within each communication regime using condition-specific models. In the raw views, later prediction error rose across both personal-anchor and group-anchor bins (Fig.~\ref{fig:bias-individual}(e),(f)). In joint models, the shared group anchor predicted later prediction error in every condition, but it outweighed the personal anchor most clearly in comprehension noise (\(p=0.0065\); control \(p=0.097\); production noise \(p=0.074\); SI Section 7.5 and SI Table~S39). A similar condition pattern appeared for later signed bias (SI Section 7.5 and SI Table~S39).

Because both anchors were measured before social information was shown, this does not mean that comprehension noise created the group anchor. Rather, comprehension noise made the already-existing shared anchor more consequential for subsequent accuracy and signed bias, while partly weakening idiosyncratic personal anchors. Other outcomes were more selective: social-signal inconsistency was most group-anchor dominated in control and production noise, and social disagreement only in production noise. The full outcome-by-condition pattern, including payoff and bias-correction models, is reported in SI Section 7.5 and SI Table~S39.

\subsubsection{\texorpdfstring{Personal anchors lowered confidence, whereas shared group anchors raised it under comprehension noise}{Personal anchors lowered confidence, whereas shared group anchors raised it under comprehension noise}}
\label{sec:anchor-confidence}

Confidence showed a particularly revealing version of this source distinction. In comprehension noise, stronger personal anchors predicted lower confidence (\(\beta=-0.183\), \(p=0.0011\)), whereas stronger shared group anchors predicted higher confidence (\(\beta=0.211\), \(p=0.00056\); coefficient-difference \(p=0.00038\); SI Section 7.5 and SI Table~S39). This opposite-sign pattern did not appear in control or production noise. Thus, under comprehension noise, personal and shared group anchors seemed to carry different subjective meanings: a strong personal anchor may have registered as uncertainty about one's own estimate, whereas a strong group anchor may have made the collective state feel more coherent. This is why Eq.~\ref{eq:confidence-state} later includes \(A_i\) and \(A_g\) separately: the same amount of initial bias can feel different when it is private to the individual versus shared by the group.

\subsubsection{\texorpdfstring{Average peer anchors provide a leave-one-out check on shared-anchor effects}{Average peer anchors provide a leave-one-out check on shared-anchor effects}}

The preceding analyses asked how a bias already present in the four-person group predicted later behavior. We next asked a closely related but distinct question: whether the bias in a participant's peers, considered separately from that participant's own first response, also predicted later individual behavior. To do this, we repeated the shared-anchor analyses from Section~\ref{sec:shared-personal-anchors} and the anchor-confidence analyses from Section~\ref{sec:anchor-confidence} with a leave-one-out social-anchor measure, defined as the absolute value of the mean signed round-1 bias of the other three group members. We call this the average peer anchor because it captures the peer context surrounding the participant rather than the full collective starting state. This analysis does not replace the shared group anchor; it tests whether the social component of that anchor remains informative when the focal participant's own initial response is removed.

In pooled cluster-robust models, the average peer anchor predicted later prediction error (\(\beta=0.486\), \(p=6.7\times10^{-5}\)), later signed bias (\(\beta=-0.455\), \(p=0.00066\)), social-signal inconsistency (\(\beta=0.325\), \(p=0.00088\)), social disagreement (\(\beta=0.307\), \(p=0.0014\)), payoff (\(\beta=-0.224\), \(p=1.6\times10^{-6}\)), and weaker bias correction (\(\beta=-0.346\), \(p=0.00013\)). Thus, the peer context alone carried information about later behavior. Descriptively, peer-anchor coefficients were slightly larger than personal-anchor coefficients for prediction error and later signed bias, but the two sources were statistically comparable rather than separable (coefficient-difference \(p=0.380\) and \(p=0.524\); SI Fig.~S24(a); SI Table~S40). This is the key contrast with the shared group anchor: peer bias mattered, but it did not dominate personal bias in the way the full group anchor did.

The condition-specific analyses led to the same interpretation and are reported in the SI (Section 7.6, SI Fig.~S24(b),(c), and SI Table~S40).

\subsubsection{\texorpdfstring{Comprehension noise corrected bias fastest over time}{Comprehension noise corrected bias fastest over time}}

Signed bias changed differently across the three communication regimes, with the fastest correction under comprehension noise (Fig.~\ref{fig:bias-individual}(g)). Because a few extreme typed responses produced visible jumps in raw round-level means, Fig.~\ref{fig:bias-individual}(g) plots median signed bias; the raw mean, outlier-excluded mean, and outlier audit are reported in SI Section 7.8, SI Fig.~S27, and SI Tables~S44--S45. The window-level tests reported below apply the same outlier rule. Averaging rounds 2--5, mean signed bias was most negative in control (\(-15.20^\circ\)), least negative in comprehension noise (\(-10.62^\circ\)), and intermediate in production noise (\(-13.42^\circ\)); control differed from comprehension noise (\(p=0.017\), Mann--Whitney test on participant averages over rounds 2--5), and comprehension noise differed from production noise (\(p=2.3\times10^{-4}\)). By rounds 20--25, all conditions remained negatively biased, but comprehension noise still showed the smallest bias (\(-9.21^\circ\)), followed by production noise (\(-10.52^\circ\)) and control (\(-12.36^\circ\)); the comprehension-production contrast remained significant (\(p=0.0011\)). Section~\ref{sec:dynamic-updates} later clarifies the mechanism behind this pattern: noisy channels kept the social-information weight, \(\sigma_i(t)\), active over time, but continued updating was most helpful when comprehension noise kept the social signal correctable rather than turning it into a shared production-noise record.

\subsubsection{\texorpdfstring{Comprehension noise stabilized weak anchors, but differences narrowed for strong anchors}{Comprehension noise stabilized weak anchors, but differences narrowed for strong anchors}}
\label{sec:weak-anchor-correction}

The anchor-range analysis shows where this advantage came from (Fig.~\ref{fig:bias-individual}(h)). Bias correction is measured relative to the round-1 anchor: positive values mean movement back toward the room temperatures, whereas negative values mean that estimates moved even farther below the environment than the participant's first response. The clearest treatment difference occurred when anchors were weak. At a representative \(5^\circ\) anchor bias, model-implied correction was essentially flat in comprehension noise (\(-0.02^\circ\)), but negative in control (\(-4.84^\circ\); \(p=0.017\) vs comprehension noise) and production noise (\(-5.05^\circ\); \(p=0.0015\) vs comprehension noise; HC3 OLS contrasts). Thus, weak downward anchors were stabilized or slightly improved under comprehension noise, but tended to worsen under control and production noise. As anchors became larger, the treatment differences became less reliable: production-comprehension contrasts were not significant at representative \(15^\circ\) and \(25^\circ\) anchors (\(p=0.590\) and \(p=0.500\)). The steeper fitted production-noise line therefore means that production noise narrowed its early disadvantage as anchors strengthened; it does not mean that production noise outperformed comprehension noise for strongly biased participants.

Together, these results suggest that comprehension noise was most beneficial when initial anchors were still weak enough to be corrected, whereas production noise was most vulnerable when those weak anchors could be converted into shared social evidence. This is the anchor-side version of the consensus result in Section~\ref{sec:harmful-consensus}: weakly biased groups were still recoverable, but production noise could turn a recoverable starting point into a more persistent wrong social state. Model details and the binned descriptive version are provided in SI Section 7.8, SI Table~S42, and SI Fig.~S25.

\subsection{Harmful consensus is most likely under production noise}
\label{sec:harmful-consensus}

The example trajectories in Fig.~\ref{fig:design}(c),(d) introduce the central consensus problem on which we focus in this section. In every round, participants saw the true temperature, so agreement could in principle have remained tied to the environment. Yet the examples show two different outcomes: one group converged toward the true temperature, whereas the other moved toward a shared but inaccurate estimate despite repeatedly available correct evidence. This resembles classical conformity in a continuous-estimation setting: evidence is present, but social information can organize responses around a wrong value \cite{asch1951,lorenz2011}. We use informed consensus for agreement that remains coupled to the true temperature, and harmful or misinformed consensus for agreement around an inaccurate value (Fig.~\ref{fig:consensus}). In the language of Eq.~\ref{eq:dynamic-update}, misinformed consensus is the collective state in which peer pull, \(S_i(t)-x_i(t)\), has aligned group members with one another, but evidence pull, \(T_t-x_i(t)\), has not recoupled the group mean to \(T_t\).

Theoretical work on collective sensing predicts that noise can sometimes help groups escape misinformed consensus \cite{salahshour2019prl}, while work on biological communication and collective signaling predicts that its effect should depend on where noise enters the communication channel \cite{salahshour2019sr,Salahshour2021frontiers}. This section therefore tests the group-level consequence of the model rather than only the individual update weights: do the different forms of \(S_i(t)\) change how long groups remain in a wrong but socially aligned state? The analyses below show two related results. First, the time-course analysis shows that harmful consensus tended to decline over time in the noisy conditions, suggesting that noisy social information can help loosen some wrong consensus states, consistent with theoretical work \cite{salahshour2019prl}. Second, production noise nevertheless made harmful or misinformed consensus more likely than comprehension noise, especially early in the interaction and among low-anchor groups. The anchor analysis explains this asymmetry: comprehension noise kept weak pre-social anchors more correctable, whereas production noise could convert those same weak anchors into shared social evidence.
\subsubsection{\texorpdfstring{Participant-level harmful consensus declined under noise}{Participant-level harmful consensus declined under noise}}

We defined participant-level harmful consensus as a state in which a participant's estimate was, on average, close to peers (\(D\leq5^\circ\)) but far from the true temperature (\(E_{\mathrm{truth}}\geq5^\circ\)). Here \(D\) is the participant's mean absolute distance from contemporaneous peer estimates. This thresholded measure is intentionally interpretable: it identifies participants who were near their peers but far from the environment. The \(5^\circ\) threshold was chosen because it matches the maximum absolute social-channel perturbation in the noisy conditions. A round-level analysis showed that the treatment difference was already visible by round 5 and reappeared at rounds 10, 20, and 25 (Fig.~\ref{fig:consensus}(a.i); SI Table~S7), indicating that comprehension noise was associated with better correction away from harmful consensus over time. Using each participant's across-round average, harmful consensus occurred in \(8/196=4.1\%\) of control observations, \(6/204=2.9\%\) of comprehension-noise observations, and \(15/200=7.5\%\) of production-noise observations (Fig.~\ref{fig:consensus}(a.ii)). The comprehension-production contrast was significant (\(p=0.039\), two-proportion z-test; SI Table~S10).

The time course also clarifies why noise can be protective. After round 5, harmful consensus declined in both noisy conditions, with a significant negative trend in production noise (\(\beta=-0.0035\) per round, \(p=0.037\), OLS trend test across sampled rounds) and a similar but weaker trend in comprehension noise (\(\beta=-0.0021\), \(p=0.070\); SI Table~S9). The control condition showed no evidence of a decline (\(\beta=-0.00014\), \(p=0.935\)). Thus, noisy social information did not simply impair performance; once participants had experienced the reliability structure of the task, noise appears to have helped loosen some wrong consensus states by motivating continued updating. The dynamic analysis in Section~\ref{sec:dynamic-updates} gives this interpretation a process-level form: noisy channels kept \(\sigma_i(t)\) positive after the earliest rounds, and comprehension noise showed the clearest evidence that this responsiveness helped groups leave an existing misinformed-consensus state.

However, production noise was particularly detrimental in the initial rounds. Harmful consensus occurred in \(13\%\) of control observations, \(10.8\%\) of comprehension-noise observations, and \(21.5\%\) of production-noise observations; the control-production contrast was significant (\(p=0.021\), two-proportion z-test), as was the comprehension-production contrast (\(p=0.003\); SI Table~S7). This shows that correlated broadcast of the same corrupted information can strongly seed harmful consensus early in the interaction.

Additional robustness analyses using a continuous measure of whether consensus remained accurate produced the same comprehension-production pattern under broad parameter choices; see the SI for details (SI Fig.~S9; SI Table~S11). The corresponding threshold sweep for the group-level duration analysis is reported below, where the group-level outcome is introduced.

\subsubsection{\texorpdfstring{Production noise increased time spent in group-level misinformed consensus}{Production noise increased time spent in group-level misinformed consensus}}
\label{sec:consensus-duration}

The participant-level analysis asks whether an individual was close to peers while far from the truth. The group-level analysis asks the parallel collective question: did the group as a whole cluster around a wrong value in a given round? We defined a group-level misinformed-consensus event as any round in which the group's mean absolute deviation around its own mean was at most \(5^\circ\), while the group mean remained at least \(5^\circ\) from the true temperature. Thus, a group-round was counted only when the submitted estimates were clustered within one possible social-noise amplitude and the group mean was at least one such amplitude away from the room temperature.

This distinction sets the evidential hierarchy. Because the central claim concerns collective information acquisition, the primary behavioral outcome is the proportion of rounds that submitted estimates spent in group-level misinformed consensus. The participant-level harmful-consensus analysis above is the individual-level counterpart: it asks whether the same agreement-with-error pattern was visible when the unit was the participant rather than the group.

The main group-level outcome is duration: the proportion of task rounds that submitted estimates spent clustered around error. Duration distinguishes brief, one-round clustering from persistent misinformed consensus, which is the collective state most relevant to the theory. We also report any-entry summaries as secondary checks in SI Fig.~S8. These summaries ask whether a group entered misinformed consensus at least once; because each group had many opportunities to enter, they can collapse transient entries and persistent states into the same binary outcome. SI Fig.~S8 also reports definition checks linking the participant-level and group-level rules.

\begin{figure}[p]
 \centering
 \includegraphics[width=\textwidth]{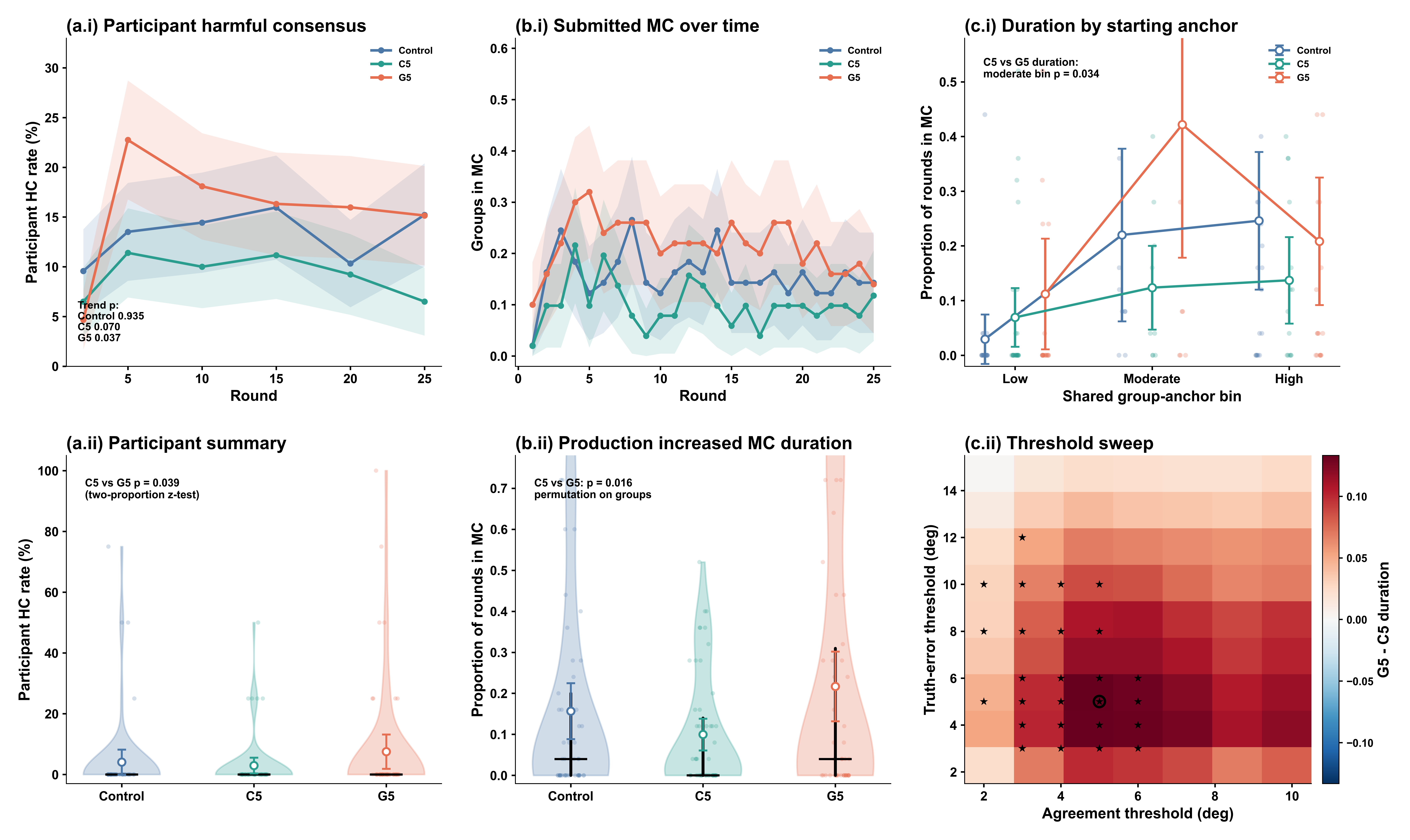}
 \caption{Harmful and misinformed consensus. Panel (a.i) shows participant-level harmful-consensus (HC) rates across sampled rounds, where HC means that a participant was close to peers (\(D\leq5^\circ\)) but far from the true room temperature (\(E_{\mathrm{truth}}\geq5^\circ\)). Panel (a.ii) summarizes the same participant-level outcome by condition using group-level participant-HC rates, shown with transparent violin shapes, group points, means, and confidence intervals. Panel (b.i) shows the proportion of groups in submitted-estimate misinformed consensus (MC) across rounds; MC is defined as a group mean absolute deviation of at most \(5^\circ\) together with a group mean at least \(5^\circ\) from the true room temperature. Panel (b.ii) summarizes the primary behavioral outcome: the proportion of rounds each group spent in submitted-estimate MC. Panel (c.i) shows the same duration outcome within shared group-anchor bins. Panel (c.ii) repeats the production-minus-comprehension duration contrast across agreement and truth-error thresholds; warmer colors indicate more time in MC under production noise, stars indicate permutation \(p<0.05\), and the circled marker denotes the primary \(5^\circ/5^\circ\) threshold. Any-entry summaries and shared-social-record versions are reported in SI Fig.~S8 and SI Tables~S10 and~S15.}
 \label{fig:consensus}
 \label{fig:submitted-consensus-duration}
\end{figure}

The group-round time series showed the same qualitative asymmetry as the participant-level analysis (Fig.~\ref{fig:submitted-consensus-duration}(b.i)). Production-noise groups began with the highest misinformed-consensus rate and showed a marginal decline after round 5 (\(\beta=-0.0079\) per round, \(p=0.068\), OLS trend test), whereas the control and comprehension-noise trends were not significant (control \(\beta=0.0020\), \(p=0.114\); comprehension noise \(\beta=0.0012\), \(p=0.477\); SI Table~S9). Using submitted estimates, production-noise groups spent a larger proportion of rounds in misinformed consensus than comprehension-noise groups (mean proportion \(=0.217\) vs \(0.100\); group-level permutation \(p=0.016\); Fig.~\ref{fig:submitted-consensus-duration}(b.ii)). The control-production contrast was weaker (mean proportion \(=0.162\) vs \(0.217\); \(p=0.278\)).

The duration result remained positive in a covariate-adjusted group-level model that controlled for room temperature, shared group-anchor magnitude, and their interaction (G5 vs C5 coefficient \(=0.110\), \(p=0.020\), HC3 OLS; SI Section 5). It was also stable across a dense threshold sweep: production noise exceeded comprehension noise descriptively across all tested agreement and truth-error thresholds, with the chosen \(5^\circ/5^\circ\) threshold lying in a supported region (Fig.~\ref{fig:submitted-consensus-duration}(c.ii); SI Section 5). This covariate-adjusted result connects the consensus analysis back to Section~\ref{sec:initial-bias}: shared anchors helped identify which groups were vulnerable, and production noise still increased misinformed-consensus duration after differences in room temperature and starting anchor structure were taken into account. The primary group-round definition used available rounds with at least three valid submissions; a complete four-person-round sensitivity preserved the duration contrast, though more conservatively (G5--C5 difference \(=0.104\), permutation \(p=0.030\); SI Section 5). Median onset, defined among groups that entered at least once as the first round in which a group entered misinformed consensus, was early and similar across conditions: round 3 in control and round 4 in both noisy conditions. Thus, the treatment difference was not that production-noise groups first entered much earlier. It was persistence: production-noise groups spent more rounds clustered around a wrong value (SI Table~S15).

We also checked the any-round statistic, which asks whether a group entered misinformed consensus at least once during the task. This measure was less sensitive to the treatment difference because each group contributed many rounds: a brief one-round entry was enough to count, and transient entries were collapsed with persistent consensus on error. It is therefore less diagnostic of the mechanism than duration: any-round misinformed consensus occurred in \(26/49=53.1\%\) of control groups, \(25/51=49.0\%\) of comprehension-noise groups, and \(27/50=54.0\%\) of production-noise groups, with no reliable comprehension-production entry contrast (\(p=0.692\), Fisher's exact test; SI Fig.~S8 and SI Table~S10).

The submitted-estimate analysis describes what groups chose. To examine the channel mechanism more directly, we also applied the same group-level misinformed-consensus rule to the shared social record. By shared social record we mean the social-channel values available to participants through \(S_i(t)\) before they made their next estimate; in production noise, this record can include a perturbed value stored before broadcast and therefore made available as common social information. This channel-level outcome asks whether the social input itself became wrong and aligned, before asking whether participants' submitted estimates remained there.

This channel-level outcome showed the same production-comprehension direction more strongly: production-noise records entered misinformed consensus more often than comprehension-noise records (\(36/50=72.0\%\) vs \(25/51=49.0\%\), \(p=0.025\), Fisher's exact test) and spent more rounds in that state (mean proportion \(=0.222\) vs \(0.100\), permutation \(p=0.004\); SI Fig.~S8 and SI Table~S15). When shared-record duration was restricted to rounds 1--24, the records that could influence a later update, the production-comprehension contrast remained positive (available-round permutation \(p=0.0025\); complete-round permutation \(p=0.0049\); SI Section 5). The direction was therefore aligned across submitted behavior and the social record: production noise exceeded comprehension noise in both. The strongest statistic differed because the two records capture different stages of the process. Shared-record entry measures whether the social input became wrong and aligned; submitted-estimate duration measures whether groups then stayed around wrong values after combining that social input with the thermometer evidence, \(T_t\), prior estimates, and uncertainty.

\subsubsection{\texorpdfstring{Anchors shaped how long groups remained in misinformed consensus}{Anchors shaped how long groups remained in misinformed consensus}}
\label{sec:anchor-consensus}

A look at the anchor structure helps explain when groups remained in misinformed consensus (Fig.~\ref{fig:submitted-consensus-duration}(c.i)). For submitted estimates, the clearest anchor-bin result was about persistence. Moderate-anchor groups spent more time in misinformed consensus under production noise than under comprehension noise (permutation \(p=0.034\)). Low-anchor groups showed the same direction but not reliably, and high-anchor groups were vulnerable in all regimes. For the shared social record, the clearest result was about entry. In low-anchor groups, production-noise records entered misinformed consensus more often than the corresponding control and comprehension-noise records (SI Section 5; SI Table~S19). These are different but compatible signals. The shared record is the social input available through \(S_i(t)\); it can become wrong and aligned before participants respond. Submitted estimates are the behavioral response to that input; they show whether groups stayed near the wrong value after combining social information with the thermometer evidence, \(T_t\), and their own prior estimates. Thus, the direction was similar across records, but the strongest statistic differed: entry for the shared record, duration for submitted behavior.

The peer-anchor robustness check gave the same message in a simpler way. We replaced the full shared group anchor with the average peer anchor, a leave-one-out measure based only on the other group members. This peer-context measure again suggested that anchors mattered for time spent in misinformed consensus, but it did not reproduce the low-anchor entry effect in the shared social record (SI Section 7.6, SI Fig.~S24(e), and SI Table~S20). This is why we treat the two anchors separately: the shared group anchor describes the group's collective starting state, whereas the average peer anchor describes the focal participant's surrounding peer context.

A second robustness check reached the same conclusion from the participant-level side. Recall that harmful consensus classified a participant-round as socially aligned but wrong: the participant's estimate was close to peers while still far from the true temperature. We applied that participant-level rule first and then summarized the resulting classifications within groups (SI Fig.~S12; SI Table~S18). The low-anchor production excess again appeared. The interpretation is straightforward. Strongly anchored groups were already at risk in every condition. Weakly anchored groups were more recoverable, but production noise could place a wrong value into the shared social record and make that value socially available to the group. Comprehension noise, by contrast, more often left distortions receiver-specific and therefore easier to escape. 

The next subsection tests this process directly at the round-to-round level: did participants continue to move toward peer pull, \(S_i(t)-x_i(t)\), and did group-level channel-noise components predict subsequent movement of the group mean?

\subsection{\texorpdfstring{Dynamic updates linked shared noise to later movement}{Dynamic updates linked shared noise to later movement}}
\label{sec:dynamic-updates}

Sections~\ref{sec:initial-bias} and \ref{sec:harmful-consensus} established two parts of the collective-error process. First, many participants entered the task with downward pre-social anchors, and the anchor already shared within a group was the stronger predictor of later error. Second, production-noise groups spent more time clustered around wrong submitted estimates, while the shared social record showed how production-noise perturbations could create consensus on error. The remaining question is how these anchor and consensus patterns emerged from one round to the next. In Eq.~\ref{eq:dynamic-update}, the social-information weight \(\sigma_i(t)\) links the social signal \(S_i(t)\) to the next estimate: when \(\sigma_i(t)>0\), participants move toward peer pull, \(S_i(t)-x_i(t)\). If production noise were more harmful because it made participants assign a larger social-information weight to peers, the production-noise increment in Eq.~\ref{eq:social-weight} should be positive, \(\sigma_P>0\). If, instead, production noise was more harmful because the same social-information weight acted on a production-noise perturbation in \(S_i(t)\), then comprehension and production noise could have similar social-information weights, \(\sigma_i(t)\), while producing different collective outcomes. This section tests these links directly: peer updating in Eq.~\ref{eq:dynamic-update} (Prediction 3), the production-specific social-information-weight increment \(\sigma_P\) in Eq.~\ref{eq:social-weight} (Prediction 4), the moderator terms in Eq.~\ref{eq:social-weight} and the confidence-state terms in Eq.~\ref{eq:confidence-state} (Predictions 7--9 and 11), and the movement of the group mean predicted by group-level channel-noise components (Prediction 6; Table~\ref{tab:theory-predictions}).

\subsubsection{\texorpdfstring{Estimating personal-evidence and social-information weights, \(\omega_i(t)\) and \(\sigma_i(t)\)}{Estimating personal-evidence and social-information weights}}

The dynamic models estimate the two coupling weights in Eq.~\ref{eq:dynamic-update}: the personal-evidence weight \(\omega_i(t)\), estimated as the coefficient on evidence pull, \(T_t-x_i(t)\), and the social-information weight \(\sigma_i(t)\), estimated as the coefficient on peer pull, \(S_i(t)-x_i(t)\). For each participant and each social round, the outcome was the change from the previous estimate to the next estimate. Predictors described the information available before that next estimate: the evidence pull \(T_t-x_i(t)\), the peer pull \(S_i(t)-x_i(t)\), and peer dispersion, \(D_i(t)\). To prevent a few extreme typed responses from dominating the slopes, the dynamic models used the same \(0^\circ\)--\(300^\circ\) round-level outlier rule as the signed-bias sensitivity analysis (SI Section 12; SI Tables~S67--S70).

The update weights make the social mechanism visible (Fig.~\ref{fig:dynamic-update}(a); SI Table~S67). Participants moved reliably toward the social signal in every condition, giving a positive social-information weight, \(\sigma_i(t)\): control \(\widehat{\sigma}=0.266\), \(p=4.1\times10^{-5}\); comprehension noise \(\widehat{\sigma}=0.507\), \(p=3.6\times10^{-15}\); production noise \(\widehat{\sigma}=0.502\), \(p=9.2\times10^{-15}\) (cluster-robust OLS). This confirms Prediction 3: participants used the social signal, \(S_i(t)\), to adjust their next estimates. The personal-evidence weight \(\omega_i(t)\), estimated as the coefficient on evidence pull, \(T_t-x_i(t)\), was small after the previous estimate and peer signal were included (SI Table~S67), indicating that later updates were more strongly tied to the social record than to direct recoupling with the thermometer cue.

The comparison between comprehension and production noise is central. Prediction 4 asks whether noisy communication changes the social-information weight, \(\sigma_i(t)\), and especially whether production noise makes \(\sigma_i(t)\) larger than comprehension noise. It did not. In the condition-only interaction model (with comprehension-noise and production-noise indicators as the only peer-pull moderators), the comprehension-control increment estimated the noisy-channel term, \(\widehat{\sigma}_N=0.240\) (\(p=0.0082\)). The production-comprehension increment estimated the production-specific term, \(\widehat{\sigma}_P=-0.004\) (\(p=0.963\)), so the total production-control increment was \(\widehat{\sigma}_N+\widehat{\sigma}_P=0.236\) (\(p=0.0097\); Fig.~\ref{fig:dynamic-update}(a); SI Table~S72). Thus, production noise was not more dangerous because participants gave peers a larger mean social-information weight than they did in comprehension noise. Rather, a similar \(\sigma_i(t)\) acted on different social signals, \(S_i(t)\), in the two noisy regimes.

\subsubsection{\texorpdfstring{Noisy channels kept social information influential, but only comprehension noise helped responsiveness to evidence}{Noisy channels kept social information influential, but only comprehension noise helped responsiveness to evidence}}

Prediction 4 is not simply about whether people followed peers. Prediction 3 already showed that they did. Prediction 4 asks what noise did to the social-information weight, \(\sigma_i(t)\), across the repeated rounds of social exchange. The clearest message is simple: in the control condition, the pull of the social signal weakened over time, whereas in both noisy conditions it remained positive (Fig.~\ref{fig:dynamic-update}(b); SI Table~S69). Noise therefore did not merely scramble communication. It kept peer information active after the earliest rounds.

Sustained \(\sigma_i(t)\) was useful only when the social signal remained correctable. At the participant level, peer pull, \(S_i(t)-x_i(t)\), that pointed back toward the evidence, \(T_t\), predicted later error reduction in both noisy conditions, including late rounds (comprehension noise: \(\beta=0.264\), \(p=0.0099\); production noise: \(\beta=0.309\), \(p=0.0003\); SI Table~S81). At the group level, however, the clearer escape signature belonged to comprehension noise: it made groups more likely than control to leave an existing misinformed-consensus state (logit \(\beta=1.038\), \(p=0.0037\); \(\beta=0.760\), \(p=0.032\) after accounting for peer dispersion, \(D_i(t)\)), whereas production noise did not (SI Table~S82). Thus, both noisy conditions kept social information influential compared with control, but continued responsiveness helped most clearly when the noise was comprehension noise. Production noise also kept \(\sigma_i(t)>0\), but the same responsiveness could act on a more correlated production-noise record. See SI Section 12 and SI Tables~S81--S82 for details.

\subsubsection{\texorpdfstring{Testing moderators of the social-information weight (\(\sigma_i(t)\))}{Testing moderators of the social-information weight}}

Equation~\ref{eq:dynamic-update} is the core dynamic model: it asks how evidence pull, \(T_t-x_i(t)\), and peer pull, \(S_i(t)-x_i(t)\), move the next estimate. Equation~\ref{eq:social-weight} then asks which factors change the social-information weight, \(\sigma_i(t)\), within that update. We tested these candidate moderators by entering personal anchors, \(A_i\), shared group anchors, \(A_g\), confidence, \(C_i(t)\), peer dispersion, \(D_i(t)\), and channel indicators, \(I_{\mathrm{noise}}\) and \(I_{\mathrm{prod}}\), as interactions with peer pull, \(S_i(t)-x_i(t)\) (SI Fig.~S45; SI Section 12.1). This adjusted model asks a different question from the condition-only model above. The condition-only model estimates the total difference in \(\sigma_i(t)\) associated with the noisy regimes. The adjusted model asks whether a residual channel effect remains after measured reliability and starting-state variables are included. In that adjusted model, the production-specific residual term, \(\sigma_P\), estimated by \([S_i(t)-x_i(t)]\times I_{\mathrm{prod}}\), was positive but not reliable (\(\widehat{\sigma}_P=0.076\), \(p=0.303\)). The residual noisy-channel term, \(\sigma_N\), estimated by \([S_i(t)-x_i(t)]\times I_{\mathrm{noise}}\), was also small and not reliable (\(\widehat{\sigma}_N=0.044\), \(p=0.600\)). Thus, the fuller model does not overturn the condition-only finding that noisy regimes showed stronger social updating than control; it indicates that this average noisy-channel increment was partly carried by confidence, \(C_i(t)\), peer dispersion, \(D_i(t)\), personal anchors, \(A_i\), and shared group anchors, \(A_g\). The central production-noise interpretation is unchanged: production was not worse because it produced a larger residual \(\sigma_i(t)\) than comprehension noise, but because a similar social-information weight acted on a more correlated production-noise social signal.

Because production noise did not increase the average social-information weight, \(\sigma_i(t)\), beyond comprehension noise, the next question was what made \(\sigma_i(t)\) stronger or weaker within the interaction. Higher previous confidence, \(C_i(t)\), reliably reduced the social-information weight, \(\sigma_i(t)\), in the pooled model (\(\widehat{\sigma}_C=-0.109\), \(p=1.5\times10^{-5}\)) and within the two noisy regimes, supporting Prediction 7. Peer dispersion, \(D_i(t)\), also attenuated the social-information weight, \(\sigma_i(t)\), in the pooled model (\(\widehat{\sigma}_D=-0.063\), \(p=0.044\)), but this moderation was strongest in production noise and should be interpreted cautiously outside that regime. Shared group anchors, \(A_g\), increased the social-information weight, \(\sigma_i(t)\), (\(\widehat{\sigma}_G=0.051\), \(p=0.00043\)), whereas personal anchors, \(A_i\), did not (\(\widehat{\sigma}_a=0.000009\), \(p=0.9996\); shared-minus-personal moderation \(p=0.063\); SI Section 12.1). This gives the shared-anchor result in Section~\ref{sec:shared-personal-anchors} a dynamic interpretation. Earlier, \(A_g\) predicted later error and signed bias more strongly than \(A_i\); here, \(A_g\), but not \(A_i\), was associated with a larger social-information weight, \(\sigma_i(t)\). The shared starting bias therefore appears to have mattered not only because it marked biased groups, but because it identified groups in which peer pull, \(S_i(t)-x_i(t)\), carried more weight during updating.

\subsubsection{\texorpdfstring{Testing predictors of confidence (\(C_i(t)\))}{Testing predictors of confidence}}

Equation~\ref{eq:confidence-state} asks whether measured confidence, \(C_i(t)\), tracked the reliability cues that theory proposes can gate social updating. The pooled model supported several of these links (SI Fig.~S45; SI Section 12.1). Confidence was lower in noisy social channels (\(\widehat{\kappa}_N=-0.552\), \(p=0.015\)), lower when peer dispersion, \(D_i(t)\), was larger (\(\widehat{\kappa}_D=-0.278\), \(p=0.043\)), and lower for participants with stronger personal anchors, \(A_i\) (\(\widehat{\kappa}_a=-0.223\), \(p=0.049\)). Current error, \(|x_i(t)-T_t|\), was negative but marginal (\(\widehat{\kappa}_E=-0.276\), \(p=0.091\)). Shared group anchors, \(A_g\), showed the opposite sign: they were positively associated with confidence in the pooled model (\(\widehat{\kappa}_g=0.324\), \(p=0.0074\)), driven most clearly by comprehension noise (\(\widehat{\kappa}_g=0.949\), \(p=4.5\times10^{-16}\)). This joint pattern connects three parts of the paper. First, it reproduces the condition effect in Section~\ref{sec:individual-effects}: noisy social channels lowered confidence. Second, it gives the anchor-confidence result in Section~\ref{sec:anchor-confidence} a dynamic interpretation: personal anchors behaved like private uncertainty, whereas shared group anchors could make the social context feel coherent and increase confidence. Third, it helps interpret the later payoff analysis in Section~\ref{sec:confidence-payoff}: confidence can regulate social updating, but it is not guaranteed to track accuracy when the social record itself carries bias.

\subsubsection{\texorpdfstring{Channel-noise components moved later estimates (\(S_i(t)-S_i^{\mathrm{unnoised}}(t)\))}{Channel-noise components moved later estimates}}

The preceding result showed that production noise did not make participants assign a larger mean social-information weight, \(\sigma_i(t)\), than comprehension noise. The remaining theoretical difference must therefore lie in the structure of the social signal, \(S_i(t)\), itself. To test that mechanism, we decomposed \(S_i(t)\) following Eqs.~\ref{eq:comprehension-signal}--\ref{eq:production-signal}. The un-noised social signal, \(S_i^{\mathrm{unnoised}}(t)\), captures peers' previous estimates before added social distortion. The channel-noise component, \(S_i(t)-S_i^{\mathrm{unnoised}}(t)\), captures the perturbation introduced by comprehension noise or production noise. This decomposition asks whether the treatment difference came from a larger production-noise \(\sigma_i(t)\), which the previous subsection did not support, or from the covariance structure of the values entering \(S_i(t)\).

\begin{figure}[p]
 \centering
 \includegraphics[width=\textwidth]{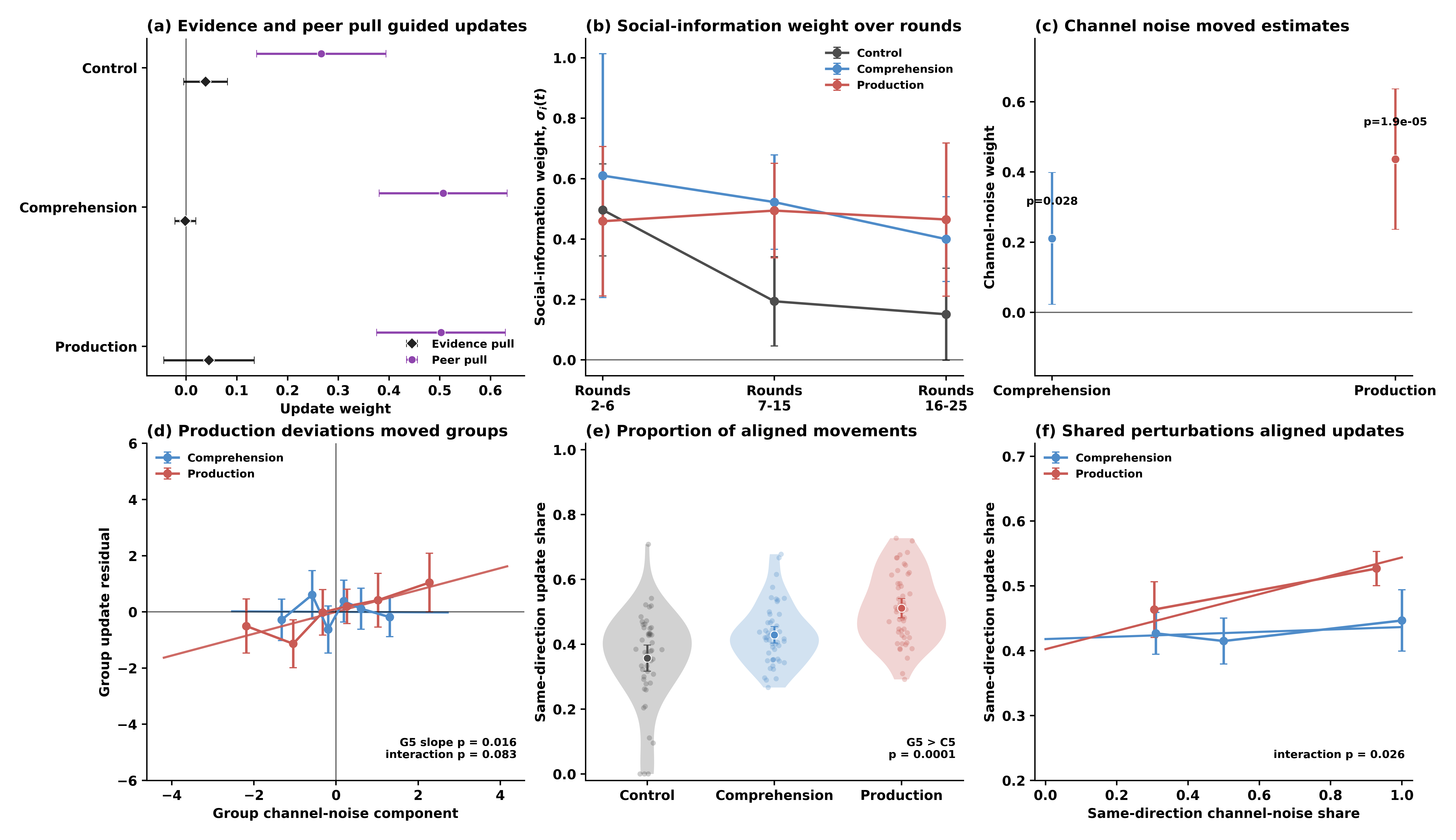}
 \caption{Dynamic update mechanism. Each observation in panels (a)--(c) is a participant-round update from rounds 2--25, defined as the next un-noised estimate minus the participant's previous un-noised estimate. Panel (a) compares the slopes on evidence pull, \(T_t-x_i(t)\), and peer pull, \(S_i(t)-x_i(t)\). These slopes estimate the personal-evidence weight, \(\omega_i(t)\), and the social-information weight, \(\sigma_i(t)\), respectively, in condition-specific cluster-robust OLS models. Panel (b) repeats the social-information weight, \(\sigma_i(t)\), across early, middle, and late rounds, showing whether the social-information weight persisted over the task. Panel (c) decomposes the social signal, \(S_i(t)\), into peers' un-noised previous estimates and the channel-noise component, \(S_i(t)-S_i^{\mathrm{unnoised}}(t)\), and asks whether the perturbation itself pulled later estimates. Panel (d) uses group-round summaries to ask whether group-level channel-noise components predicted the next movement of the group mean. Panel (e) shows the group-level same-direction update share. For each group-round, participant updates \(\Delta x_i=x_i(t+1)-x_i(t)\) were converted to signs after excluding zero or missing updates; all unordered participant pairs were coded 1 if their signs matched and 0 otherwise; and the group-round share was averaged across rounds within each group. Two groups had no defined group-round shares after zero and missing updates were omitted and therefore do not contribute to panel (e). Panel (f) applies the same pairwise sign logic to the channel-noise component and asks whether same-direction channel perturbations predicted same-direction submitted updates in the noisy conditions. Error bars are 95\% confidence intervals clustered by four-person group where applicable. Full Eq.~\ref{eq:social-weight} moderator tests, confidence-state models, covariance analyses, and behavioral synchrony models are reported in SI Section 12 and SI Fig.~S45.}
 \label{fig:dynamic-update}
\end{figure}

We used two related tests. The participant-level test asked whether each person moved toward the channel-noise component in their own social signal. They did: channel-noise pull predicted the next update in comprehension noise (\(\beta=0.211\), \(p=0.028\)) and production noise (\(\beta=0.437\), \(p=1.9\times10^{-5}\); Fig.~\ref{fig:dynamic-update}(c)). The production coefficient was larger, but the formal production-comprehension difference was only suggestive (\(p=0.104\)). This result shows that perturbed social values affected individual updating, but it is not the strongest test of the channel-location asymmetry.

The group-round test asked the more specific cascade question in Prediction 6: when perturbations were averaged at the group level, did they move the next group mean? Here the evidence was clearest for production noise. The group-level channel-noise component predicted the next movement of the group mean in production noise (\(\beta=0.389\), \(p=0.016\)), but not in comprehension noise (\(\beta=-0.009\), \(p=0.961\); production-comprehension interaction \(p=0.083\); Fig.~\ref{fig:dynamic-update}(d)). This is the dynamic counterpart of the shared-record result in Section~\ref{sec:consensus-duration}: production noise not only made the stored social record more likely to contain consensus on error, but those group-level perturbations also predicted the next movement of submitted estimates. The within-production slope was reliable (\(p=0.016\)), whereas the direct production-comprehension difference between slopes was suggestive (\(p=0.083\)). The update model therefore provides mechanistic support rather than definitive proof; the strongest treatment-asymmetry evidence remains the group-level consensus-duration and shared-record analyses.

The formal covariance check and an illustrative counterfactual are reported in SI Section 12. The observed group-mean channel-noise variance was close to the theoretical comprehension-noise value under comprehension noise (0.841 vs 0.833) and much larger under production noise (2.228 vs the theoretical production-noise value of 2.500). Receiver-to-receiver noise correlations showed the same pattern (approximately 0.004 for comprehension noise and 0.640 for production noise). In the counterfactual check, making production-noise perturbations receiver-specific moved the covariance structure toward comprehension noise, whereas making comprehension-style perturbations shared before display moved it toward production noise. The check is not a behavioral simulation, but it is a useful diagnostic of the channel mechanism: changing only whether perturbations were receiver-specific or shared was enough to move the realized social record between the two covariance regimes predicted by Eqs.~\ref{eq:comprehension-signal}--\ref{eq:production-signal}.

This covariance pattern is implied by the channel equations, but the behavioral question is whether the same structure also appeared in submitted belief movements. We therefore asked whether updates were more coordinated when the channel perturbations were more coordinated. For each group-round, we calculated the same-direction update share from participant updates \(\Delta x_i=x_i(t+1)-x_i(t)\): after omitting missing or zero updates, all unordered participant pairs were coded 1 if their \(\Delta x_i\) values had the same sign and 0 otherwise, and the group-round share was the mean of those pair codes. Figure~\ref{fig:dynamic-update}(e) plots each group's average share across social rounds for groups with at least one defined group-round share; two groups had only missing or zero update pairs under this rule and therefore did not contribute to this panel. Production-noise groups had the highest group-level same-direction update share (production \(0.511\), comprehension \(0.429\), control \(0.357\); production-comprehension \(p=0.00013\), Mann--Whitney test; Fig.~\ref{fig:dynamic-update}(e)). In the noisy conditions, the same pairwise sign calculation applied to the channel-noise component predicted same-direction submitted updates under production noise but not comprehension noise (production simple slope \(=0.141\), \(p=0.0011\); comprehension simple slope \(=0.018\), \(p=0.598\); interaction \(p=0.026\); Fig.~\ref{fig:dynamic-update}(f)). This behavioral synchrony result is not a claim that coordinated movement is always harmful. It shows that production noise did not merely create a mathematically correlated channel record; it made submitted belief movements more coordinated when the channel perturbations themselves were coordinated.

\subsubsection{\texorpdfstring{Held-out validation and parsimony of the update model}{Held-out validation and parsimony of the update model}}

The preceding analyses were mechanistic fits: they estimated whether the coupling terms in Eqs.~\ref{eq:dynamic-update}--\ref{eq:confidence-state} were present in the observed data. We next asked whether the same structure had out-of-sample predictive value. This analysis should be read as a theory-guided parsimony check, not as an exhaustive search for the statistically minimal model. Leave-one-group-out validation held out one four-person group at a time, fit the models on the remaining groups, and predicted the held-out group. The validation was one-step-ahead: it used the observed previous estimate, \(x_i(t)\), and observed held-out covariates to predict the next update, \(x_i(t+1)-x_i(t)\). All standardization was learned within the training groups before predicting the held-out group.

The validation compared three models. Model 1 was a condition-specific reference model. It retained the two update terms in Eq.~\ref{eq:dynamic-update}: evidence pull, \(T_t-x_i(t)\), and peer pull, \(S_i(t)-x_i(t)\), while allowing their slopes to differ freely by condition. Peer dispersion, \(D_i(t)\), is not included in the minimal update equation because it is not a directional target: unlike \(T_t-x_i(t)\) or \(S_i(t)-x_i(t)\), it does not specify whether the next estimate should move upward or downward. The theory therefore treats \(D_i(t)\) as a reliability cue that can change the social-information weight, \(\sigma_i(t)\), in Eq.~\ref{eq:social-weight}. Model 1 nevertheless included \(D_i(t)\) as an additive empirical control, making the reference model conservative: later gains from Eq.~\ref{eq:social-weight} then reflect moderation of \(\sigma_i(t)\), not simply giving the richer models access to peer-dispersion information.
\[
x_i(t+1)-x_i(t)=
\alpha_{c(i)}
+\omega_{c(i)}\big[T_t-x_i(t)\big]
+\sigma_{c(i)}\big[S_i(t)-x_i(t)\big]
+\delta_{c(i)}D_i(t)+\epsilon_i(t),
\]
where \(c(i)\) indexes the participant's condition. Relative to Eq.~\ref{eq:social-weight}, this reference model allows the average social-information weight, \(\sigma_{c(i)}\), to differ across control, comprehension noise, and production noise, but it sets the psychological and anchor moderation terms to zero: \(\sigma_C=\sigma_D=\sigma_G=\sigma_a=0\). Thus, Model 1 can use \(D_i(t)\) to predict updates directly, but it does not allow \(D_i(t)\) to change \(\sigma_i(t)\). Model 2 added the theory-motivated social-information-weight moderators except the personal anchor,
\[
\sigma_i^{\mathrm{shared}}(t)=
\sigma_0+\sigma_N I_{\mathrm{noise}}+\sigma_P I_{\mathrm{prod}}
+\sigma_C C_i(t)+\sigma_D D_i(t)+\sigma_G A_g,
\]
with corresponding main effects included. Model 3, the full model, then added personal-anchor moderation,
\[
\sigma_i^{\mathrm{full}}(t)=\sigma_i^{\mathrm{shared}}(t)+\sigma_a A_i.
\]
Thus the validation asked two questions. First, do the Eq.~\ref{eq:social-weight} moderators improve prediction beyond a flexible condition-specific update rule? Second, does personal-anchor moderation, \(A_i\), add predictive value once the shared group anchor, \(A_g\), is already included?

\begin{figure}[p]
 \centering
 \includegraphics[width=\textwidth]{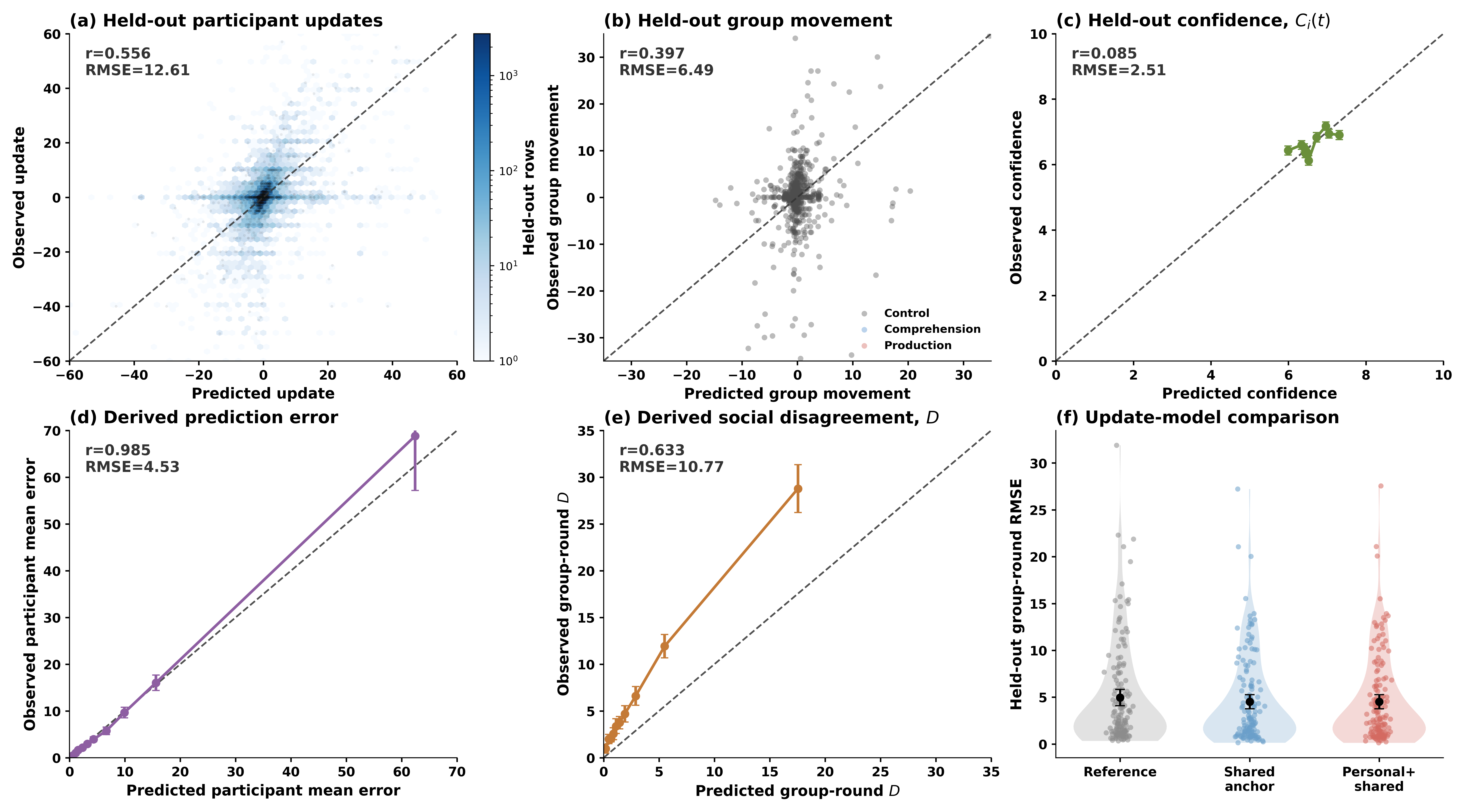}
 \caption{Leave-one-group-out validation. Each fold held out one four-person group, fit the relevant model on the remaining groups, and predicted the held-out group. The update validation is one-step-ahead: it uses the observed previous estimate, \(x_i(t)\), and observed held-out covariates to predict the next update, \(x_i(t+1)-x_i(t)\), not a free-running trajectory. Panels (a),(b) compare observed and predicted updates from the full Eq.~\ref{eq:social-weight} update model at the participant-round and group-round levels. Panel (c) validates the confidence-state equation, Eq.~\ref{eq:confidence-state}, by predicting held-out confidence, \(C_i(t)\), from observed reliability cues. Panels (d),(e) add the one-step-ahead predicted updates to observed previous estimates and use the resulting predicted estimates to derive participant-level prediction error and group-round social disagreement, \(D\). Dashed lines mark perfect prediction. Panel (f) compares held-out group-round RMSE for the condition-specific reference model, the update model with shared-anchor terms, and the full update model with both shared-anchor and personal-anchor terms; points are held-out groups, violin shapes show group distributions, and black markers show means with 95\% confidence intervals. Full comparisons are reported in SI Section 12.3.}
 \label{fig:oos-validation}
\end{figure}

The answer was partly positive and partly conservative. The full model captured a meaningful part of unseen updating: predicted and observed participant-round updates were positively correlated (\(r=0.556\); Fig.~\ref{fig:oos-validation}(a)), as were predicted and observed group-round movements (\(r=0.397\); Fig.~\ref{fig:oos-validation}(b)). Across all held-out rows, RMSE was \(12.61^\circ\) for participant-round updates and \(6.49^\circ\) for group-round movements. When RMSE was first computed within each held-out group and then averaged across groups, adding the Eq.~\ref{eq:social-weight} moderator structure reduced participant-round RMSE from \(8.86^\circ\) to \(8.47^\circ\) and group-round RMSE from \(4.96^\circ\) to \(4.52^\circ\) relative to the condition-specific reference model (Fig.~\ref{fig:oos-validation}(f); SI Section 12.3). However, adding \(A_i\) on top of the shared-anchor model did not improve group-round prediction: the shared-anchor model had slightly lower group-round RMSE than the full model (\(4.51^\circ\) vs \(4.52^\circ\)). The predictive gain therefore came mainly from the shared-anchor, confidence, peer-dispersion, and channel terms, not from personal-anchor moderation of \(\sigma_i(t)\).

We also evaluated quantities that are derived from, or parallel to, the update model rather than directly estimated as Eq.~\ref{eq:social-weight} coefficients. First, the one-step-ahead update predictions were added to the observed previous estimates and used to derive downstream behavioral summaries. These one-step-ahead predicted estimates preserved participant-level prediction-error structure well (\(r=0.985\), RMSE \(=4.53^\circ\); Fig.~\ref{fig:oos-validation}(d)) and captured group-round social disagreement, \(D\), moderately (\(r=0.633\), RMSE \(=10.77^\circ\); Fig.~\ref{fig:oos-validation}(e)), although large disagreement excursions were underpredicted. Second, Eq.~\ref{eq:confidence-state} was validated separately by predicting held-out confidence, \(C_i(t)\), from observed reliability cues. This prediction was weak (\(r=0.085\), RMSE \(=2.51\) on the 0--10 scale; Fig.~\ref{fig:oos-validation}(c)), indicating that the confidence-state equation is better treated as an explanatory association than as a strong individual-level forecasting model. Thus, the model has useful predictive power for one-step-ahead belief updates and derived accuracy/disagreement summaries, but it does not fully predict confidence or rare large jumps.

\subsection{\texorpdfstring{Exploratory extension: post-event reconstruction links social context to behavior}{Exploratory extension: post-event reconstruction links social context to behavior}}

 The preceding sections describe online behavior: participants began with anchors, updated toward the social signal, \(S_i(t)\), and sometimes converged on consensus on error. We next ask a more psychological question. After the interaction ended, did the same anchors and social records leave traces in how participants reconstructed the environment and their own behavior? This analysis is exploratory because reconstruction was measured after the main task and is not part of the primary dynamic evidence. Still, the theory gives it a natural motivation: if anchors and social records shape online estimates, they may also shape how participants later remember the environment and their own behavior.

The link to Section~\ref{sec:dynamic-updates} is therefore conceptual rather than primary evidence. The dynamic models showed that estimates were pulled by peer pull, \(S_i(t)-x_i(t)\); the reconstruction measures ask whether traces of those online estimates and social records remained visible after the task. These questions separate what participants did during the task from how they later represented what had happened. The two numerical questions used here were asked immediately after the main task and were quoted exactly as follows: ``What do you think was the average temperature during the game?'' and ``What do you think was the average of the estimate that you provided during the game?'' The first answer was compared with the objective average temperature sequence to define environment-reconstruction error, \(\Delta_{\mathrm{env}}\); the second was compared with the participant's data-derived average estimate to define self-reconstruction error, \(\Delta_{\mathrm{self}}\). The environment-self gap, \(\Delta_{\mathrm{env-self}}\), is the first recalled value minus the second. This terminology is deliberately retrospective: participants were not asked to report a known answer, but to reconstruct the task environment and their own behavior after the interaction had ended.

\subsubsection{\texorpdfstring{Environmental reconstruction was downward and less precise under production noise}{Environmental reconstruction was downward and less precise under production noise}}

Detailed distributions and tests are reported in SI Section 6.2, SI Fig.~S13, and SI Tables~S23 and~S25--S26. Absolute self-reconstruction error and the absolute environment-self gap did not show reliable condition differences. The clearest condition pattern was environmental imprecision: absolute environment-reconstruction error was larger in production noise than in comprehension noise (\(p=0.043\)). This contrast was affected by one extreme production-noise response, but the direction remained the same after excluding it and the rank-based contrast became marginal (\(p=0.052\)). We therefore treat the production-noise absolute reconstruction contrast as suggestive rather than primary evidence.

The signed reconstruction errors gave a simpler directional message. Participants did not reliably misremember their own average estimate upward or downward, but they reconstructed the average room temperature as lower than the objective average in every condition and in the pooled sample (all sign-rank tests \(p\leq5.8\times10^{-23}\); pooled \(p=3.8\times10^{-69}\)). The robust post-task pattern is therefore downward environmental reconstruction across conditions, while the production-noise absolute-error pattern indicates greater dispersion or imprecision rather than a clean directional shift.

These results suggest that production noise may have hindered participants' ability to construct a stable representation of the environmental history after the task, not only their moment-to-moment estimates during the task. The interpretation parallels the dynamic result in Section~\ref{sec:dynamic-updates}: production-noise channel-noise components in \(S_i(t)\) predicted subsequent group movement more clearly than comprehension-noise channel-noise components. The reconstruction result is less robust than the online consensus result, but it points to the same channel distinction: a comprehension-noise perturbation was specific to a receiver, whereas a production-noise perturbation could enter the common social record before participants received it.

\subsubsection{\texorpdfstring{Initial anchors leaked into remembered experience}{Initial anchors leaked into remembered experience}}

The recalled average-temperature answer tracked participants' own average and final estimates more strongly than the objective average temperature, and signed environmental reconstruction followed signed task bias during the experiment (SI Sections 6.3--6.4, SI Fig.~S15(a),(b), and SI Tables~S27--S29). This supports treating the post-task ``average temperature'' answer as a reconstructed belief about the environment rather than as an independent readout of the objective average.

Interestingly, the first-round anchor also reappeared in the post-task questionnaire. Stronger absolute round-1 pre-social anchor bias predicted larger gaps between recalled values and their corresponding benchmarks: larger absolute self-reconstruction error (\(\beta=0.091\), \(p=0.012\)), larger absolute environment-reconstruction error (\(\beta=0.102\), \(p=0.0012\)), and a more negative signed environment-reconstruction error (\(\beta=-0.129\), \(p=2.5\times10^{-5}\); SI Table~S28). The link to self-reconstruction was clearest in production noise, where stronger initial anchors predicted larger absolute self-reconstruction error (\(\beta=0.125\), \(p=0.027\); SI Table~S28). Thus, initial anchors did not merely predict what participants later reported; they predicted reconstruction error, the discrepancy between a recalled value and the corresponding data-derived or objective value.

\subsubsection{\texorpdfstring{Social instability increased memory-reality gaps}{Social instability increased memory-reality gaps}}

Social volatility, defined as the average round-to-round change in the group mean estimate, was the clearest social-context predictor of self-reconstruction (SI Fig.~S15(c); SI Table~S30). More volatile group trajectories predicted larger absolute self-reconstruction error (\(\beta=0.230\), \(p=3.4\times10^{-9}\)) and a more negative signed self-reconstruction error (\(\beta=-0.164\), \(p=8.6\times10^{-5}\)). Thus, unstable social records made participants' own behavior harder to reconstruct. This is consistent with the dynamic-update evidence in Section~\ref{sec:dynamic-updates}: peer pull, \(S_i(t)-x_i(t)\), shaped subsequent estimates, so a rapidly moving group record also created a more difficult memory target.

\subsubsection{\texorpdfstring{Memory of one's own estimates was pulled toward the group}{Memory of one's own estimates was pulled toward the group}}

Social volatility did not, however, replace self-memory with group memory: recalled own estimates were still closer to participants' actual own average than to the participant-level average social signal, \(\bar S_i\), the group average, or the actual environmental average (all paired sign-rank tests \(p\leq2.2\times10^{-5}\); SI Table~S30). Instead, self-reconstruction was modestly pulled toward the social record. When social information was higher than a participant's own average estimate, remembered own average estimates shifted upward, and when it was lower, they shifted downward (\(\beta=0.133\), \(p=0.0017\), cluster-robust OLS; SI Table~S30). This retrospective pull toward the social record is weaker and more indirect than the online social-information weight, \(\sigma_i(t)\), but it has the same sign structure: the remembered self moved in the direction of the social stream. Together, these patterns suggest that participants did not simply replay the task record after the experiment. Their retrospective reports appear to have been partly tilted toward the social stream: participants reconstructed their own behavior, and to some extent the environmental history, in a way that reduced remembered distance from the group context. The corresponding environmental-reconstruction models are reported in SI Section 6.4, SI Fig.~S15(c), and SI Table~S29.
\subsection{\texorpdfstring{Exploratory extension: confidence and social traits}{Exploratory extension: confidence and social traits}}

We next asked which variables predicted payoff, prediction error, confidence, and social-signal inconsistency. Section~\ref{sec:dynamic-updates} analysed the task dynamically, at the participant-round level: it tested how confidence, \(C_i(t)\), peer dispersion, \(D_i(t)\), and anchors changed the social-information weight, \(\sigma_i(t)\), and how channel-noise components shaped subsequent movement. The analyses here ask a different, aggregated question. After each participant's trajectory was summarized across the task, which personal and group-context variables were associated with performance, confidence, and alignment? These regressions therefore connect to the same theoretical quantities, but they do not estimate the update equation itself.

The goal was to separate the roles of anchors, confidence, peer confidence, and social orientation rather than interpret each predictor in isolation. To keep these tests comparable, we used one integrated condition-specific regression framework. The model entered personal anchor magnitude, shared group anchor magnitude, own confidence, peer confidence, susceptibility, and collectivism together; own confidence was omitted only when confidence itself was the outcome. Peer confidence was the leave-one-out mean confidence of the other group members. It was not displayed to participants, so it should be interpreted as a group-context covariate rather than as a directly observed social cue. Susceptibility indexed willingness to rely on or be influenced by others \cite{mehrabian1995}, whereas collectivism indexed broader endorsement of group-oriented values \cite{Singelis1995,Triandis1998}. These traits were included to test whether individual differences in social orientation changed how participants used social information.

Predictors and outcomes were standardized within condition, and standard errors were clustered by four-person group (Fig.~\ref{fig:regression}(a)--(c); SI Table~S47). Additional behavioral models with narrower predictor sets are reported in the SI as robustness checks (SI Tables~S48--S49 and~S53; SI Figs.~S28--S29).

\subsubsection{Confidence predicted payoff less clearly under production noise}
\label{sec:confidence-payoff}

\begin{figure}[p]
 \centering
 \includegraphics[width=\textwidth]{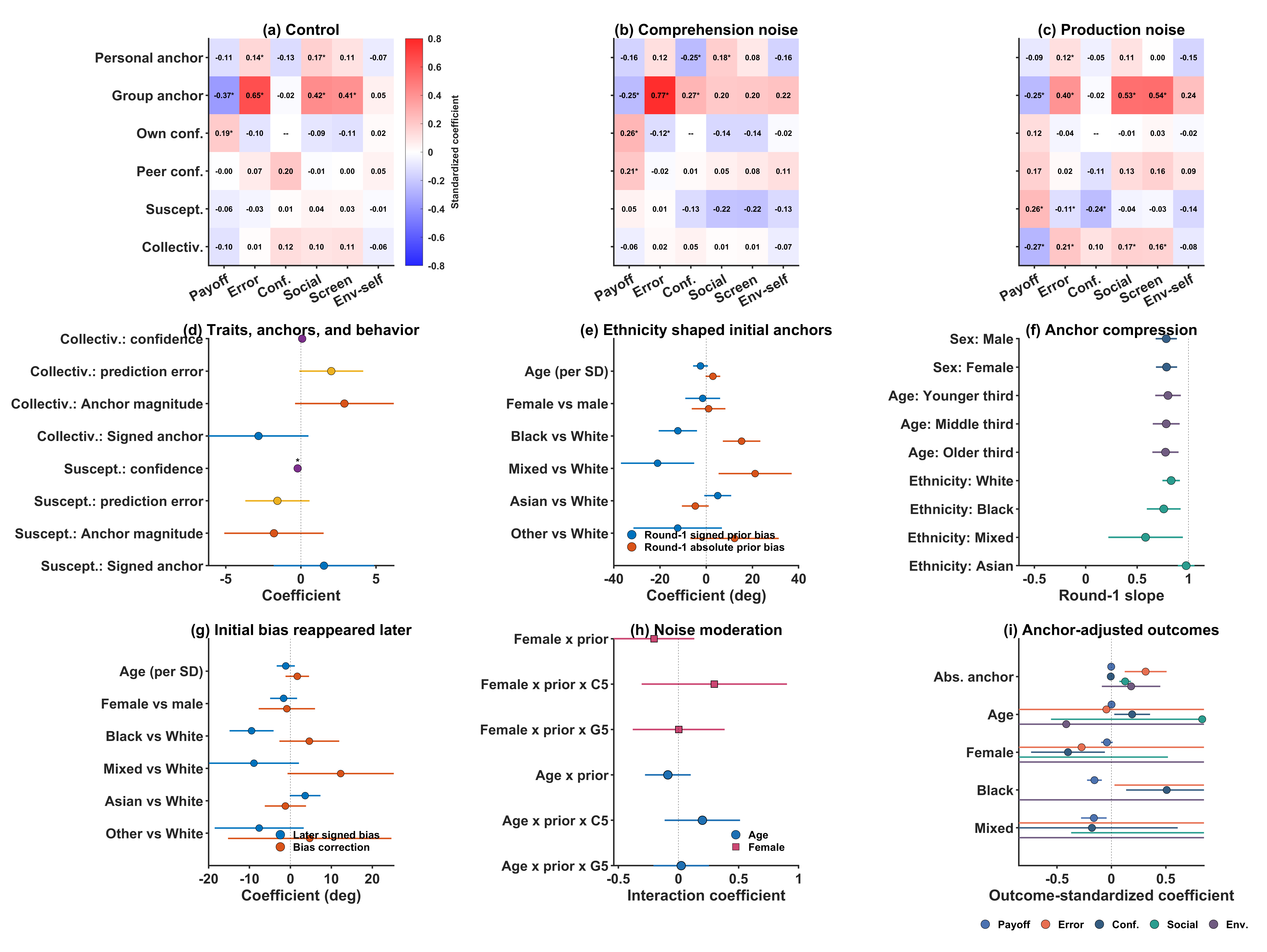}
 \caption{Individual-difference and integrated regression analyses. Panels (a)--(c) show integrated treatment-specific regressions for control, comprehension noise, and production noise. The color scale in panel (a) applies to panels (a)--(c); coefficients are standardized and white cells indicate omitted self-predictor cells. Panel (d) shows pooled survey-trait associations with signed round-1 anchor, round-1 anchor magnitude, later prediction error, and confidence. Panels (e)--(i) summarize bias-aware demographic analyses; in panel (h), colors distinguish age interactions from female-sex interactions. Asterisks mark \(p<0.05\). Full models are reported in SI Tables~S54--S57, S60--S61, and~S47; sensitivity models are shown in SI Figs.~S28--S29.}
 \label{fig:regression}
 \label{fig:demographics}
\end{figure}

Own confidence predicted higher payoff in control (\(\beta=0.188\), \(p=0.0038\); Fig.~\ref{fig:regression}(a)) and comprehension noise (\(\beta=0.264\), \(p=2.0\times10^{-6}\); Fig.~\ref{fig:regression}(b)), while the corresponding coefficient in production noise was weaker and only marginal (\(\beta=0.122\), \(p=0.058\); Fig.~\ref{fig:regression}(c)). This suggests that under production noise, participants could express high confidence without being comparably well anchored to the environment. Together with the dynamic result that confidence reduced the peer-pull weight, \(\sigma_i(t)\), this indicates that confidence had two roles: it regulated social updating, but it did not always remain a reliable marker of accuracy when the shared record could be corrupted.

Peer confidence predicted payoff in comprehension noise (\(\beta=0.205\), \(p=0.011\)), but not in control (\(\beta=-0.00003\), \(p=0.9997\)) and only marginally in production noise (\(\beta=0.167\), \(p=0.072\)). Because participants did not see peers' confidence ratings, this result means that being embedded in a more confident peer context covaried most clearly with higher payoff under comprehension noise. We tested this interpretation directly by linking peer confidence to the direction of peer pull. For each social round, round-specific peer confidence was the mean confidence of the peers whose estimates formed \(S_i(t)\), and evidence-aligned peer pull was \([S_i(t)-x_i(t)]\,\mathrm{sign}[T_t-x_i(t)]\), which is positive when the social signal points back toward the evidence. Peer confidence did not reliably predict a more accurate displayed social signal, \(S_i(t)\), in either noisy condition, and it did not modulate immediate round-level error reduction. At the participant level, however, peer confidence interacted with average evidence-aligned peer pull to predict payoff in comprehension noise (\(\beta=0.166\), \(p=0.025\)), but not in production noise (\(\beta=0.091\), \(p=0.357\)); the formal production-comprehension difference was not reliable (\(p=0.354\); SI Section 8.1). Thus, the data give suggestive aggregate support for the idea that confident peer contexts were most useful when peer pull remained evidence-aligned under comprehension noise, but they do not show a decisive round-by-round confidence-gating pathway.

\subsubsection{Shared group bias predicted poorer performance}

The clearest predictor of performance was the bias shared by the group. Stronger shared group anchors predicted lower payoff in all three conditions (control \(\beta=-0.365\), \(p=0.00034\); comprehension noise \(\beta=-0.248\), \(p=0.0019\); production noise \(\beta=-0.252\), \(p=0.0013\)) and higher prediction error in all three conditions (control \(\beta=0.646\), \(p=0.0013\); comprehension noise \(\beta=0.772\), \(p=4.1\times10^{-5}\); production noise \(\beta=0.397\), \(p=0.0016\)). In other words, participants embedded in more biased groups performed worse even after accounting for their own anchor, their confidence, peer confidence, susceptibility, and collectivism. This complements the bias analysis in Section~\ref{sec:initial-bias} by showing that shared group anchors remained consequential inside the fuller behavioral model. The link to Section~\ref{sec:dynamic-updates} is mechanistic rather than redundant. The shared group anchor, \(A_g\), is the absolute value of the group's mean first-round signed bias, so a larger \(A_g\) means that peers began with a common displacement from the evidence \(T_t\). Section~\ref{sec:dynamic-updates} shows that \(A_g\), but not the personal anchor \(A_i\), increased the social-information weight, \(\sigma_i(t)\), on peer pull, \(S_i(t)-x_i(t)\). Thus, shared bias predicted later error not only because it marked groups that started farther from the evidence, but because that shared starting state also made the peer signal more influential during subsequent updating.

\subsubsection{\texorpdfstring{Peer-confidence context predicted a larger remembered environment-self gap}{Peer-confidence context predicted a larger remembered environment-self gap}}

The clearest integrated-model signal for the remembered environment-self gap was peer confidence. Higher peer confidence was associated with a larger absolute gap between remembered environment and remembered own estimate in the pooled integrated model (\(\beta=0.065\), \(p=0.018\); SI Table~S47), with similar but weaker condition-specific trends in comprehension and production noise. This does not show that participants explicitly felt wrong during the task, because peer confidence was not displayed. It instead suggests a retrospective separation between two remembered quantities: in groups whose other members were generally more confident, participants later kept their remembered own estimates farther from their remembered environment. One possible reason is that confident group contexts made the social record more stable or salient, giving participants a clearer post-task reference against which their own estimates could be reconstructed. 

The corresponding susceptibility-moderation check is reported in SI Section 8.1. It did not support the stronger claim that the peer-confidence association was concentrated among highly susceptible participants.

\subsubsection{\texorpdfstring{Susceptibility helped, whereas collectivism hurt, under production noise}{Susceptibility helped, whereas collectivism hurt, under production noise}}

The effects of susceptibility and collectivism depended on where noise entered the communication channel, which is important because the same social tendency can be adaptive in one informational ecology and maladaptive in another.

The main trait result was concentrated in production noise. Susceptibility predicted higher payoff (\(\beta=0.255\), \(p=2.0\times10^{-6}\)), lower prediction error (\(\beta=-0.113\), \(p=0.021\)), and lower confidence (\(\beta=-0.242\), \(p=4.4\times10^{-5}\); Fig.~\ref{fig:regression}(c)). This pattern suggests cautious social responsiveness: susceptible participants were less confident, but they were not worse performers, and under production noise they earned more. Collectivism showed the opposite profile in the same condition, predicting lower payoff (\(\beta=-0.266\), \(p=0.00011\)), higher prediction error (\(\beta=0.213\), \(p=0.00018\)), greater social disagreement (\(\beta=0.168\), \(p=0.0035\)), and greater social-signal inconsistency (\(\beta=0.162\), \(p=0.0045\)). The contrast is informative because both traits capture social orientation, but they separated sharply here. Thus, the result is not that ``being social'' was good or bad. Flexible responsiveness to social information was beneficial when the broadcast record could be corrupted, whereas a broader group-oriented disposition was costly. Because these traits were not included as peer-pull moderators in Section~\ref{sec:dynamic-updates}, this remains an exploratory extension rather than direct evidence about \(\sigma_i(t)\).

Taken as an exploratory pattern, the contrast suggests that these social-orientation traits mattered most when the social record itself was unreliable: susceptibility looked like cautious responsiveness to social uncertainty, whereas collectivism was costly when production noise could make the group record compelling despite its corruption. The weaker condition-specific patterns outside production noise are reported in SI Section 8.1.

\subsubsection{\texorpdfstring{Susceptibility was associated with weaker initial anchors}{Susceptibility was associated with weaker initial anchors}}

A complementary trait analysis asked whether these social-orientation measures were already related to the pre-social anchor before any peer information appeared. Higher susceptibility predicted a less negative round-1 anchor (\(\beta=5.65\), \(p=0.048\)), smaller absolute anchor magnitude (\(\beta=-6.09\), \(p=0.020\)), lower later prediction error (\(\beta=-4.66\), \(p=0.0039\)), and lower confidence (\(\beta=-0.497\), \(p=0.0021\); Fig.~\ref{fig:regression}(d); SI Table~S57). Collectivism did not reliably predict anchor magnitude or bias correction. Thus, susceptibility was not only beneficial under production noise once social information appeared; it was also associated with a weaker initial anchor at the start of the task. This links the trait result back to the first stage of the model. A weaker \(A_i\) cannot identify whether susceptibility changed the expectation term \(\mu_i\), the evidence weight \(\lambda_i\), or the residual displacement \(\zeta_i\) in Eq.~\ref{eq:initial-mixture}. It does show, however, that trait heterogeneity was already visible in the evidence--expectation mixture, before the social-information weight, \(\sigma_i(t)\), could operate.

\subsection{\texorpdfstring{Exploratory extension: demographic heterogeneity}{Exploratory extension: demographic heterogeneity}}

We also ran exploratory demographic regressions on the anonymized dataset, with age, sex, simplified ethnicity, and condition entered together (SI Table~S58). In relation to the theory, these variables are possible sources of heterogeneity in expectations, anchors, confidence, and social alignment; they are not alternative manipulations of the communication channel. They therefore connect most directly to the anchor and confidence components of the model, \(A_i\), \(A_g\), and \(C_i(t)\), rather than to the channel mechanism tested in Section~\ref{sec:dynamic-updates}. These models should be interpreted cautiously because demographic categories were not experimentally balanced and some ethnicity groups were small. They may also reflect recruitment pool, language, country of residence, platform composition, or cultural interpretations of the unitless temperature scale.

We distinguish two questions: whether demographic variables were associated with behavior overall, and whether they moderated the effect of noise itself. The goal is descriptive: to document heterogeneity in anchors, confidence, and performance, and to check whether this heterogeneity changed the channel-noise contrast.

\subsubsection{\texorpdfstring{Ethnicity, not age or sex, shaped pre-social anchors}{Ethnicity, not age or sex, shaped pre-social anchors}}

We first asked whether demographic groups differed in the initial anchor itself. Age and sex were not reliable predictors of either round-1 signed anchor bias or absolute anchor magnitude (Fig.~\ref{fig:regression}(e); SI Table~S54). Ethnicity was more consequential. Relative to White participants, Black participants showed a more negative round-1 anchor (\(\beta=-12.30^\circ\), \(p=0.0037\)) and a larger absolute anchor magnitude (\(\beta=15.06^\circ\), \(p=0.0003\)), and mixed-ethnicity participants showed the same pattern (\(\beta=-21.52^\circ\), \(p=0.0081\); \(\beta=21.32^\circ\), \(p=0.0082\)).

The pre-social compression analysis led to a similar conclusion (Fig.~\ref{fig:regression}(f); SI Table~S55). The slope of round-1 prediction on actual temperature did not differ by sex (interaction \(p=0.978\)) or age (interaction \(p=0.583\)). By contrast, the mixed-ethnicity slope was flatter than the White slope (\(\Delta\beta=-0.243\), \(p=0.050\)), whereas the Asian slope was steeper (\(\Delta\beta=0.155\), \(p=0.003\)). Thus, the clearest demographic heterogeneity in pre-social compression again appeared across ethnicity rather than across age or sex.

\subsubsection{\texorpdfstring{Ethnicity-linked anchors persisted into later bias}{Ethnicity-linked anchors persisted into later bias}}

These initial differences largely reappeared later (Fig.~\ref{fig:regression}(g); SI Table~S54). Black participants retained a more negative later signed bias than White participants (\(\beta=-9.52^\circ\), \(p=0.0005\)), while mixed-ethnicity participants showed a more positive bias-correction coefficient (\(\beta=12.69^\circ\), \(p=0.057\)), suggesting greater movement away from the initial anchor despite a stronger initial anchor. Noise-by-age and noise-by-sex moderation of anchor correction was limited (Fig.~\ref{fig:regression}(h); SI Table~S56): none of the age or sex interaction terms with anchor bias and condition was reliable (all \(p\geq0.215\)). Exploratory ethnicity-specific correction interactions were more irregular and are therefore reported in SI Fig.~S21(a) and SI Table~S56.

\subsubsection{\texorpdfstring{Age and sex predicted confidence after anchor adjustment}{Age and sex predicted confidence after anchor adjustment}}

Models without anchor adjustment showed that demographic variables were related to confidence, social disagreement, and several performance measures (SI Table~S58). We therefore repeated the key demographic models after including absolute anchor magnitude. The confidence associations were robust: older participants remained more confident (\(\beta=0.191\), \(p=0.023\)), and women remained less confident (\(\beta=-0.399\), \(p=0.021\)). Black participants also still showed higher prediction error than White participants after anchor adjustment (\(\beta=4.30\), \(p=0.045\); Fig.~\ref{fig:regression}(i); SI Table~S57). By contrast, the age and sex associations with social disagreement weakened once anchor bias was included. Thus, demographic differences in confidence were not simply a by-product of different initial anchors, whereas some differences in social disagreement were partly explained by anchor bias. This matters for interpreting Eq.~\ref{eq:social-weight}: because \(C_i(t)\) reduced the social-information weight, \(\sigma_i(t)\), in Section~\ref{sec:dynamic-updates}, demographic variation in confidence may change social updating indirectly, but the demographic analyses do not show that age or sex changed the channel-specific noise effect.

\subsubsection{\texorpdfstring{Demographic moderation of noise effects was limited}{Demographic moderation of noise effects was limited}}

Finally, demographic moderation of the noise effect was limited. Age and sex did not reliably moderate condition effects on prediction error, confidence, social disagreement, reconstruction, or bias outcomes in either the unadjusted or bias-aware models. The only suggestive exception was payoff: after anchor adjustment, women tended to benefit less from comprehension noise (\(\beta=-0.294\), \(p=0.092\); SI Tables~S59 and~S60--S61). Thus, within the measured demographic variables, the main channel-noise pattern was not concentrated in a single age or sex subgroup.

\subsection{\texorpdfstring{Boundary condition: GPT agents treated noise differently from human groups}{Boundary condition: GPT agents treated noise differently from human groups}}

The GPT experiment tested how language-model agents respond to noisy social communication, and whether comprehension and production noise create the same asymmetry observed in humans. In theory terms, the GPT agents received the same evidence \(T_t\), social signals \(S_i(t)\), and channel regimes, but could differ in the anchor process, confidence response, and social-information-weight dynamics. The comparison therefore asks whether the human pattern from Section~\ref{sec:dynamic-updates}--positive social-information weight without a production-specific increase in \(\sigma_i(t)\), together with vulnerability to production-noise channel components--also appears in artificial agents. While secondary to the human experiment, this comparison addresses an important question for AI-mediated communication: whether artificial agents respond to corrupted social information in the same way as human groups. Because GPT agents do not share the same embodied temperature expectations, incentives, attention limits, or memory constraints as human participants, this comparison serves as a boundary condition rather than a direct psychological model of the human data. GPT-3.5-turbo and GPT-4o-mini were chosen to compare an earlier widely used chat model with a newer lightweight model. Sampling temperature 0 probes deterministic response policies, while temperature 1 allows stochastic variation.

\subsubsection{\texorpdfstring{Noise changed GPT prediction error without reliably improving near-ceiling accuracy}{Noise changed GPT prediction error without reliably improving near-ceiling accuracy}}

\begin{figure}[p]
 \centering
 \includegraphics[width=\textwidth]{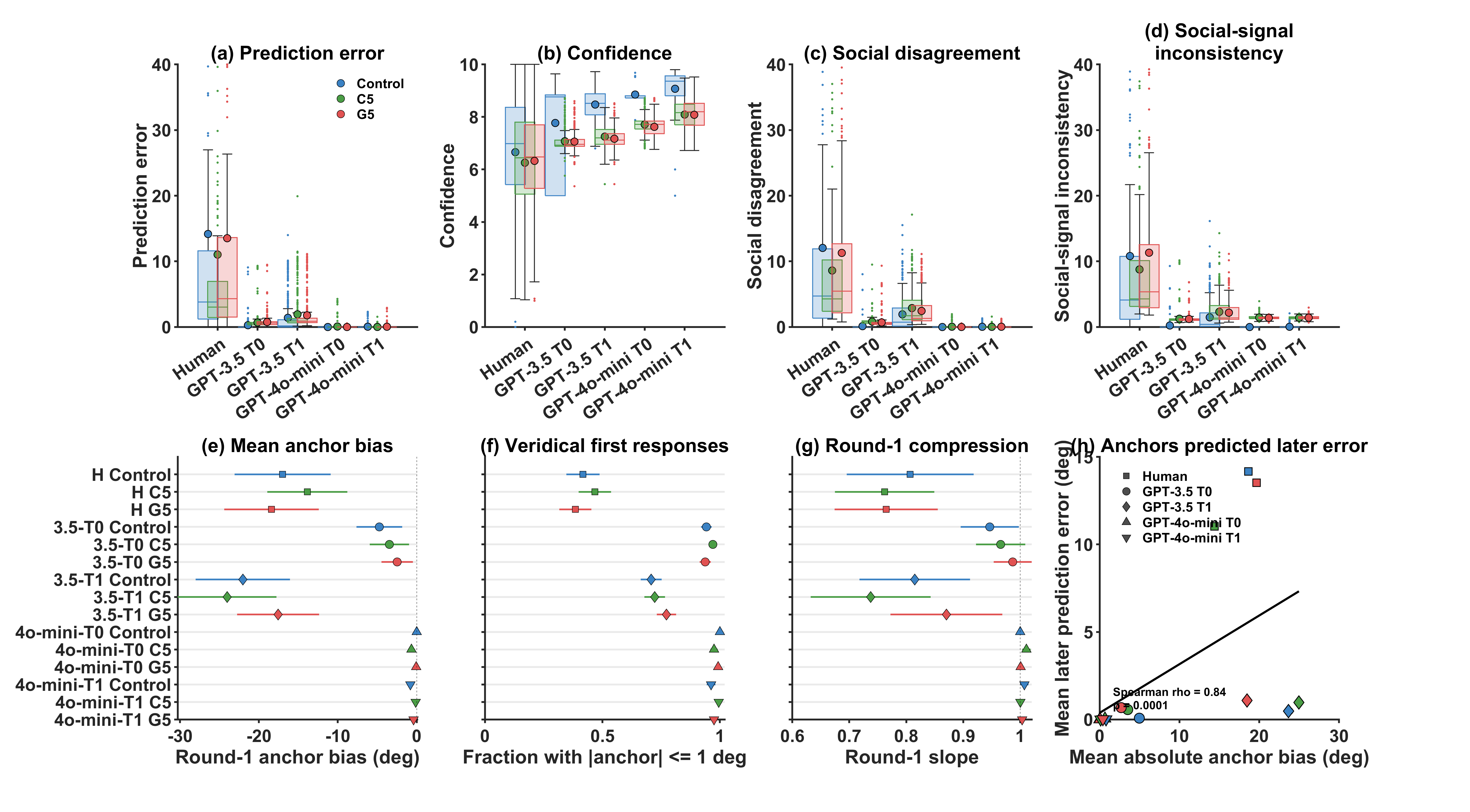}
 \caption{Human and GPT-agent comparison. Panels (a)--(d) show human and GPT-agent distributions for prediction error, confidence, social disagreement, and social-signal inconsistency, with condition indicated by color. For both humans and GPT agents, this inconsistency measure is computed relative to social-signal values. Panels (a),(c),(d) are truncated at \(40^\circ\) for readability; full-range versions are shown in SI Fig.~S35. Panels (e)--(h) add the anchor-bias comparison: mean round-1 pre-social anchor bias, the fraction of nearly veridical round-1 responses (\(|\mathrm{anchor\ bias}|\leq1^\circ\)), round-1 compression slopes, and the cross-system relation between mean absolute anchor bias and later prediction error. Panels (e)--(g) use marker shape for system and color for condition; panel (h) uses color for condition, marker shape for system, and a black regression line to show the cross-system Spearman relation. The agent-level anchor-bias regressions are shown in SI Fig.~S34. Full GPT pairwise tests and bias summaries are reported in SI Tables~S62 and~S63--S64.}
 \label{fig:gpt-human}
 \label{fig:gpt-bias}
\end{figure}

Noisy social information did not have a single effect on GPT accuracy. GPT-3.5-turbo was the more noise-sensitive model. At sampling temperature 0, both noisy conditions increased prediction error relative to control, and production noise increased prediction error more than comprehension noise (Fig.~\ref{fig:gpt-human}(a); both control-noise contrasts \(p<10^{-96}\), comprehension-production \(p=0.0023\), Mann--Whitney tests). At sampling temperature 1, GPT-3.5-turbo again became less accurate in both noisy conditions than in control (both \(p<10^{-29}\)), but the comprehension-production difference disappeared (\(p=0.793\)).

GPT-4o-mini showed a different pattern. Its prediction errors were near zero in most model-temperature-condition combinations, and the small treatment differences did not form a stable human-like asymmetry. At sampling temperature 0, comprehension noise produced more error than production noise (\(p=0.015\)), reversing the human direction. At sampling temperature 1, comprehension and production noise did not differ (\(p=0.530\)); production noise was statistically lower than control (\(p=0.028\)), but the absolute difference was only \(0.007^\circ\). Thus, noise sometimes changed GPT prediction error, but it did not reliably improve already near-ceiling GPT accuracy or reproduce the human production-noise vulnerability. Full model-specific tests are reported in SI Section 10.1 and SI Table~S62.

\subsubsection{\texorpdfstring{Noisy social information consistently reduced GPT confidence}{Noisy social information consistently reduced GPT confidence}}

Confidence was the more stable point of convergence between humans and GPT agents. Across all four model-temperature combinations, both noisy conditions reduced confidence relative to control (Fig.~\ref{fig:gpt-human}(b); all control-noise contrasts \(p<10^{-23}\), Mann--Whitney tests). The models therefore registered noisy social input as uncertainty even when their answers remained highly accurate. This resembles the human confidence-state result in Eq.~\ref{eq:confidence-state}, where noisy social channels reduced confidence, but it did not lead to the same human-scale production-noise cascade.

\subsubsection{\texorpdfstring{Humans showed stronger production-noise vulnerability than GPT agents}{Humans showed stronger production-noise vulnerability than GPT agents}}

The human-GPT comparison therefore separates uncertainty tracking from human-like collective vulnerability. Humans showed much larger prediction errors than either GPT model, as well as larger social disagreement and social-signal inconsistency (Fig.~\ref{fig:gpt-human}(a),(c),(d)). Humans also showed the clearer comprehension-production asymmetry and the stronger harmful-consensus pattern, whereas GPT agents often retained near-ceiling accuracy, especially GPT-4o-mini. We do not treat GPT post-task reconstruction as a human-like memory comparison, because GPT agents had persistent within-run histories and could answer the post-task questions from the recorded task context; those descriptive checks are reported in the SI as record-access diagnostics. Thus, GPT agents treated noisy social information as an uncertainty cue, but they did not show the same human-scale vulnerability through which shared production errors distorted collective belief formation, bias correction, and retrospective understanding. The boundary condition is therefore precise: reducing confidence under noisy social input is not sufficient to reproduce the human mechanism. The human vulnerability also required anchors and social records that could become consequential through repeated updating.

\subsubsection{\texorpdfstring{GPT anchors were downward when they appeared}{GPT anchors were downward when they appeared}}

GPT anchors add a second layer to this comparison. When GPT agents showed a pre-social anchor bias, it was downward rather than upward (Fig.~\ref{fig:gpt-human}(e)--(g)). This mirrors the human direction but not the human distribution. GPT-4o-mini was almost unbiased at round 1: mean anchor bias ranged from \(0.00^\circ\) to \(-0.81^\circ\), every median anchor bias was \(0^\circ\), and \(96.2\%\) to \(100\%\) of round-1 responses lay within \(1^\circ\) of the true temperature. GPT-3.5-turbo at sampling temperature 0 also showed only mild negative anchor bias (means \(-2.48^\circ\) to \(-4.74^\circ\), medians \(0^\circ\), near-veridical rate \(93.8\%\) to \(97.0\%\)). The strongest GPT anchor bias appeared in GPT-3.5-turbo at sampling temperature 1, where mean round-1 bias ranged from \(-17.57^\circ\) to \(-24.04^\circ\) even though the median anchor bias remained \(0^\circ\). Thus, the GPT bias was usually much weaker and sparser than the human bias, except in GPT-3.5-turbo at temperature 1, where a minority of large downward misses pulled the mean into the human range.

\subsubsection{\texorpdfstring{Weak GPT anchors helped accuracy, but did not fully explain the human-GPT gap}{Weak GPT anchors helped accuracy, but did not fully explain the human-GPT gap}}

The anchor analysis explains part, but not all, of the GPT-human difference. Across the 15 human and GPT system-condition combinations, larger mean absolute round-1 pre-social anchor bias predicted larger mean later prediction error (\(\rho=0.84\), \(p=9.1\times10^{-5}\), Spearman; Fig.~\ref{fig:gpt-human}(h)). Within GPT configurations, larger absolute anchor bias predicted more negative later signed bias in GPT-3.5-turbo at sampling temperature 0 and in GPT-4o-mini at both sampling temperatures (all \(p\leq0.0092\); SI Fig.~S34(a); SI Table~S64). Larger anchor bias also predicted higher later prediction error in the two GPT-4o-mini configurations (both \(p<3\times10^{-5}\); SI Fig.~S34(b); SI Table~S64). GPT-4o-mini's near-zero anchor therefore helps explain its high accuracy. But anchor bias is not sufficient: GPT-3.5-turbo at sampling temperature 1 showed much more human-like negative mean anchors while still remaining far more accurate than humans. This contrast echoes the human results in Sections~\ref{sec:initial-bias} and \ref{sec:dynamic-updates}: anchors matter, but they become most consequential when the later social record gives them a route into repeated updating. Detailed bias summaries are reported in SI Section 10.2 and SI Fig.~S34 and SI Tables~S63--S64.

\subsection{\texorpdfstring{Mechanistic synthesis}{Mechanistic synthesis}}

Taken together, the Results support a staged account of collective information acquisition. The first stage is pre-social: estimates were already displaced from the evidence \(T_t\), and the shared group anchor, \(A_g\), predicted later error more strongly than the personal anchor, \(A_i\). The second stage is collective: production noise increased the time groups spent in submitted-estimate misinformed consensus, while the shared social record showed how a noised value could become common evidence. The third stage is dynamic: participants moved toward peer pull, \(S_i(t)-x_i(t)\), in all regimes, but the production-specific social-information-weight increment, \(\sigma_P\), was essentially zero. Thus, production noise was not most dangerous because participants gave peers a larger mean social-information weight; it was dangerous because a similar \(\sigma_i(t)\) acted on a more correlated production-noise record.

The exploratory sections place this mechanism in a broader psychological context without changing the evidential hierarchy. Reconstruction results suggest that anchors and social records shaped later memory, but these effects are secondary and partly outlier-sensitive. Confidence and trait analyses suggest that \(C_i(t)\), \(A_g\), and social orientation changed how participants used or benefited from social information, but only confidence and \(A_g\) were directly tested as dynamic social-weight terms. Demographic analyses documented heterogeneity in anchors and confidence but did not explain away the channel effect. GPT agents provided a boundary condition: they registered noisy social information as uncertainty, yet did not reproduce the human transition from downward anchors to production-noise consensus on error. The core mechanism is therefore narrow enough to be testable and broad enough to organize the additional findings: noisy communication matters because it changes whether social evidence remains correctable or becomes a shared record that can stabilize error.

\section{Discussion}

Our study shows that noise, and especially the location of noise in the communication channel, can shape collective information acquisition. Participants repeatedly had access to a veridical thermometer cue, but the cue was subjectively uncertain and often conflicted with everyday expectations about plausible room temperatures. In that setting, communication noise did not have a single effect. Comprehension noise, which perturbed the social signal separately for each receiver, was associated with lower prediction error than production noise and sometimes helped loosen mistaken convergence. Production noise, which perturbed values before storage or display, increased the time groups spent clustered around inaccurate submitted estimates and produced the clearest consensus-on-error signature in the shared social record. The result supports collective-sensing theory, in which moderate noise can prevent maladaptive convergence, and communication theory, in which comprehension and production noise differ because they create different social-signal covariance structures \cite{salahshour2019prl,salahshour2019sr,Salahshour2021frontiers}.

This interpretation depends on keeping the evidential hierarchy explicit. Participant-level prediction error was informative and pointed in the same direction, but the most distinctive evidence was collective: production-noise groups spent more time in submitted-estimate misinformed consensus, and the shared social record showed how production-noise perturbations could create common evidence for a wrong value. These outcomes are complementary rather than interchangeable. Submitted estimates describe what groups actually chose; the shared record describes the channel-level evidence available for later social updating. Production noise, therefore, increased the persistence and social availability of collective error.

The dynamic update model clarifies the mechanism. Participants used social information in all regimes: the social-information weight, \(\sigma_i(t)\), was positive and was similar in comprehension and production noise. The production-noise effect therefore was not mainly a change in average social weighting. It came from the evidence on which that weighting acted. Comprehension-noise perturbations were almost uncorrelated across receivers, so they could be averaged out, contradicted, or discounted. Production-noise perturbations were correlated because the corrupted value entered the shared record; when participants updated toward the social signal, \(S_i(t)-x_i(t)\), this shared perturbation could move the collective state. The dynamic interaction tests were not all decisive, so the strongest treatment evidence remains the submitted-estimate consensus-duration result and the shared-record analyses. The update model nevertheless gives a process-level account of how production-noise perturbations can become consequential.

The same distinction matters because the correlated production-noise record was not only a feature of the display. That correlation was expected by design: once a production-noise perturbation was stored, several receivers could see the same corrupted value. The behavioral result is that this correlated error also entered belief updates. Production-noise channel components predicted later movement of the group mean, and production-noise groups showed more same-direction belief updates. Thus, production noise did not merely place a correlated error in the shared social record; it helped move submitted estimates together toward that erroneous signal.

These findings refine the usual interpretation of consensus. Wisdom-of-crowds arguments emphasize the value of combining partially independent information, whereas social-influence work shows that interaction can make errors correlated and reduce useful diversity \cite{galton1907,surowiecki2004,hong2004,asch1951,lorenz2011,moussaid2013,jayles2017}. The harmful- and misinformed-consensus analyses separate these two possibilities. Participants sometimes agreed with one another while remaining far from the true temperature, even though the correct cue was repeatedly available. The problem was therefore not social influence itself, but social influence that became uncoupled from evidence and was reinforced because it was socially common \cite{asch1951,gigone1993,yaniv2007,becker2017}. This is also why the time course is informative: harmful consensus tended to decline in the noisy conditions but increased in control, consistent with the idea that moderate perturbation can keep groups updating rather than locking into an early wrong attractor \cite{salahshour2019prl,flache2017}.

Pre-social anchors explain why the task was difficult before communication began. Participants appeared to compress high thermometer readings downward, plausibly because the phrase ``room temperature'', the absence of a named unit, and the warning that the thermometer could be noisy made high values seem uncertain. The data support a broad anchor account rather than a simple unit-label explanation: country-level variation was substantial, but the small Fahrenheit-major subgroup was not less biased, and round-1 anchors predicted later bias at both participant and group levels. This anchor structure matters for the noise effect. Comprehension noise was most protective when anchors were weak or moderate and still correctable; when anchors were strong, all communication regimes became harder to separate. The comparison between personal and shared group anchors further shows that the collective starting state was not reducible to individual bias. Shared group anchors predicted later error more strongly, whereas personal anchors were more closely tied to confidence, suggesting that what the group carried forward shaped accuracy while what each participant personally brought into the task shaped subjective certainty.

This anchor result is important because it turns an apparent task peculiarity into part of the mechanism. The experiment did not ask people to average arbitrary numbers; it asked them to judge a state under cue uncertainty, prior expectations, and social feedback. In many real collective-information settings, the same structure appears: people receive evidence, decide whether it is plausible, and then observe how others respond. The risk is that a shared initial displacement from evidence becomes socially reinforced before it has been corrected.

The secondary analyses describe psychological conditions under which the same mechanism was more or less harmful. Confidence predicted payoff in control and comprehension noise, but not in production noise, indicating that confidence was informative when the social record remained faithful or receiver-specific, and less diagnostic when corrupted values could be broadcast to the group \cite{bahrami2010,koriat2012}. Social disagreement predicted error in all conditions and was strongest under production noise, showing that disagreement was not simply healthy diversity but often reflected failure to coordinate around the evidence. Trait and demographic analyses add useful heterogeneity but should be interpreted cautiously because they are observational. Susceptibility and collectivism behaved differently under production noise, suggesting that flexible responsiveness to social information and broader group orientation can have different consequences when the shared record may be corrupted \cite{mehrabian1995,Singelis1995,Triandis1998}. Age and sex were related to confidence and disagreement, but they did not reliably moderate the main noise effect; demographic variation therefore does not explain away the central channel-level asymmetry. Together, these findings show that the channel mechanism operates in a psychological system where confidence, group orientation, and retrospective reconstruction matter, while remaining secondary to the group-level consensus evidence.

Post-task reconstruction extended the psychological interpretation. Reconstruction remained downward and initial anchors leaked into remembered experience, suggesting that noisy social environments can affect retrospective belief as well as online estimation \cite{loftus1974,lewandowsky2012}. GPT agents provided a boundary condition. Like humans, they reduced confidence under noisy social information, but they remained far more accurate, especially GPT-4o-mini. Their weak anchors help explain this dissociation, but not completely: GPT-3.5-turbo at sampling temperature 1 showed more human-like downward anchors while still avoiding human-scale production-noise vulnerability. Thus, registering noisy social information as uncertainty is not sufficient to reproduce the human mechanism. The human pattern required anchors and social records that could become consequential through repeated updating.

This boundary condition is useful for AI-mediated communication. AI systems increasingly summarize, store, recommend, and rebroadcast social information; if they alter where noise enters the channel, they may change collective outcomes even when users or agents appear locally responsive to uncertainty \cite{argyle2023,aher2023,park2023,herd2025,bringingeveryone2025}.

Overall, noisy communication does not simply help or hurt collective intelligence. Its effect depends on where the perturbation enters the channel. Receiver-side comprehension noise can keep biased social evidence private enough to remain correctable; sender-side production noise can turn perturbations into common evidence and stabilize consensus on error. This supports theoretical work on collective sensing and communication asymmetry, and identifies confidence, anchors, social orientation, and retrospective reconstruction as pathways through which noisy social information becomes consequential \cite{salahshour2019prl,salahshour2019sr,Salahshour2021frontiers}.

\section{Methods}

\subsection{Human experiment}

\subsubsection{Participants and ethics}

The human experiment was conducted in June--July 2024 and received ethical approval from the Ethics Committee of the University of Konstanz (IRB statement 29/2023). Participants were recruited online via Prolific, provided informed consent before participation, and completed the study in oTree. The analyzed dataset contained 600 participants: 196 in the control condition, 204 in the comprehension-noise condition (C5), and 200 in the production-noise condition (G5). We linked the behavioral data to Prolific demographic exports for 599 of the 600 analyzed participants. These rows represented 597 unique Prolific identifiers (IDs), because two participants appeared twice in the matched demographic export. Among the matched participants, 582 reported age (range 18--66 years; mean \(M=29.75\), standard deviation (SD) \(=8.57\), median \(=27\)). Reported sex was female for 299 participants, male for 279 participants, prefer not to say for 1 participant, and expired, revoked, or missing for 20 participants. After excluding expired, revoked, or missing residence entries, the largest countries of residence were South Africa (176), Poland (78), Portugal (72), the United Kingdom (30), Spain (25), Mexico (18), Italy (18), Greece (13), Hungary (13), and Canada (13). Descriptive country summaries included countries of residence with at least eight matched participants, adding Germany (11), France (8), Kenya (8), and the United States (8). Country-level models that also used country-level covariates, such as country mean susceptibility and collectivism, required at least eight participants with nonmissing values for those covariates. Under this complete-covariate rule, the United States contributed only seven complete cases and was therefore excluded from the complete-covariate country tables.

\subsubsection{Experimental design and task}

The experiment was designed as a repeated collective-sensing problem in which participants had to combine their own thermometer cue with social information from peers. Participants were randomly assigned to four-person groups and completed a repeated temperature-estimation task over 25 rounds (49 groups in the control condition, 51 groups in the comprehension-noise (C5) condition, and 50 groups in the production-noise (G5) condition). At the start of each group's task, the room temperature (the task temperature) was drawn uniformly at random between 50 and 250; this room temperature then remained fixed for that group across all 25 rounds. Participants observed a personal thermometer cue, reported a temperature estimate, and rated confidence on a 0--10 scale. From round 2 onward, they also saw social information derived from other group members' estimates in the immediately preceding round. Peer confidence ratings were recorded but were not displayed to participants; the social information consisted of peers' previous estimates. Participants were told that the thermometer showed the temperature of a room and that the thermometer could be noisy. Objectively, in the implemented experiment the displayed thermometer value matched the room temperature in every round, and the experimental manipulation affected only the social channel. The task therefore did not add experimental noise to the personal thermometer display. It tested how participants resolved a situation in which the available cue was objectively veridical but subjectively uncertain, because both the reliability of the thermometer and the social information were uncertain from the participant's perspective.

The numerical temperature scale was intentionally left without a conventional unit label. The interface did not state Celsius, Fahrenheit, Kelvin, or any other real-world scale. We retain the degree symbol in figures and statistics as shorthand for task-scale temperature units, but the experimental values should be read as unitless room-temperature values rather than conventional Celsius or Fahrenheit degrees. Because the room temperature was drawn uniformly at random from 50 to 250, values in the upper part of this range were high relative to ordinary room-temperature expectations in everyday temperature scales. We refer to the first-round estimate as the pre-social anchor. It is related to prior belief, but it is not a pure prior: it was made before any peer information appeared, but after the participant had seen the thermometer. It could therefore reflect prior expectations about room temperature, perceived cue reliability, uncertainty about the intended scale, and interpretation of the task. We define pre-social anchor bias as the pre-social anchor minus the displayed room temperature; because the thermometer display equaled the room temperature in the analyzed sessions, negative values indicate downward bias. Table~\ref{tab:variables} summarizes the main variables used in the analyses.

\subsubsection{Noise manipulation}

The key manipulation concerned where noise entered the social channel (Fig.~\ref{fig:design}). Participants in all conditions were explicitly told that the information they received about other participants' beliefs could be noisy. In the control condition, social information was displayed without added social distortion. In the comprehension-noise condition (C5), each participant's submitted estimate was stored correctly, but each receiver viewed an independently perturbed version of that previous estimate. In the production-noise condition (G5), a participant's submitted estimate was perturbed before storage and transmission, so the perturbed value could become common social evidence for multiple receivers. In both noisy conditions the perturbation was an integer drawn uniformly from \([-5,5]\). 

Formally, production noise was implemented by first storing a noisy version of each submitted estimate,
\[
\tilde b_i(t)=b_i(t)+\eta_i^{G}(t),
\]
where \(b_i(t)\) is participant \(i\)'s estimate and \(\eta_i^{G}(t)\sim U[-G,G]\). Comprehension noise was implemented at the point of perception:
\[
s_{ij}(t)=\tilde b_j(t-1)+\eta_{ij}^{C}(t),
\]
where \(\eta_{ij}^{C}(t)\sim U[-C,C]\) is independently sampled for receiver \(i\) viewing sender \(j\)'s previous estimate. 

\subsubsection{Incentives and post-task measures}

Each session lasted about 20 minutes. In addition to a fixed participation payment (about $2$ British Pounds), participants received a performance-dependent bonus (mean approximately 0.5 British Pounds). Payoff was based on the un-noised submitted estimate \(\hat T_t\): each round contributed \(1/(|\hat T_t-T_t|+1)\) when \(|\hat T_t-T_t|<10^\circ\), and 0 otherwise, yielding the cumulative payoff
\[
P=\frac{5}{100}\sum_{t=1}^{25}\frac{\mathbf{1}(|\hat T_t-T_t|<10^\circ)}{|\hat T_t-T_t|+1}.
\]
Thus, a perfect estimate contributed 0.05 British Pounds to the bonus for that round, a \(1^\circ\) error contributed 0.025 British Pounds, a \(5^\circ\) error contributed about 0.008 British Pounds, and errors of \(10^\circ\) or more contributed nothing.
The game page did not display round-by-round accuracy or payoff feedback; participants were told the rule and the accumulated bonus was computed from their submitted estimates.

After the estimation task, participants completed a custom post-task questionnaire about the remembered temperature environment, their own average estimate, and their perceived deviation from the room temperature. These post-task items were designed to measure retrospective reconstruction: how they later represented their own performance and the environment. Participants then completed two social-orientation questionnaires. The first was the 11-item Mehrabian and Stefl conformity scale, which measures susceptibility to interpersonal influence, such as relying on advice, yielding in disagreement, and aligning with others' views \cite{mehrabian1995}. The second was an adapted horizontal/vertical individualism--collectivism item set derived from Singelis et al. and Triandis and Gelfand, including items on self-reliance, competition, cooperation, respect for group decisions, and family obligation \cite{Singelis1995,Triandis1998}. These questionnaires captured both willingness to rely on social information and broader social orientation.

\subsubsection{Analysis strategy}

Analyses were organized by the level of the claim. Unless otherwise stated, group-level misinformed-consensus analyses used all available group-rounds with at least three valid estimates; complete four-person group-round sensitivity checks and the shared-record 1--24-round window check are reported in the SI. Individual-level outcomes were summarized at the participant level, using each participant's average across the 25 rounds unless otherwise noted. All behavioral accuracy, bias, anchor, and payoff analyses used participants' un-noised submitted estimates. Analyses of the shared social record are labelled separately because, in production noise, the stored value could differ from the participant's own submission. Collective outcomes were analysed at the group, group-round, or cluster-robust level. The primary collective tests were the group-level contrasts comparing comprehension and production noise on misinformed consensus: whether a group entered the state and how many rounds it spent there. Participant-level prediction error, confidence, and social-signal inconsistency were treated as convergent individual-level summaries rather than as the sole basis for collective claims. Condition contrasts were tested with two-sided Mann--Whitney tests. Because participants were nested within four-person groups, the same pairwise contrasts were also checked with group means and cluster-robust ordinary least-squares (OLS) models using the four-person groups as the clustering unit. We also added group-level permutation tests for the main condition contrasts, and we report duration and onset summaries for group-level misinformed consensus. Harmful-consensus rates were compared with two-proportion z-tests. OLS regressions were fit on complete cases for the variables included in each model; in standardized regressions, predictors and outcomes were z-scored before fitting so coefficient magnitudes could be compared on a common scale. Post-task reconstruction, survey-trait, demographic, and GPT-agent analyses are presented as exploratory or boundary-condition analyses that help interpret the primary group-level pattern.

\subsection{GPT-agent experiment}

The GPT experiment was designed as a secondary boundary-condition analysis rather than a substitute for the human experiment. It used the same task framing and social-channel conditions to ask how GPT agents behave under comprehension and production noise, and how those responses compare with human groups. The analysed GPT dataset contained 100 four-agent groups for each model, sampling-temperature, and condition combination (400 agents per cell), crossing GPT-3.5-turbo and GPT-4o-mini with sampling temperatures 0 and 1. Agents had persistent within-run message histories, responses were parsed in the required ``temperature, confidence'' format, and peer confidence was not displayed. For GPT experiments, for each model, sampling temperature, and condition, per-round predictions, confidence reports, actual temperatures, and displayed values were summarized into agent-level metrics analogous to the human participant-level measures. As in the human experiment, the environmental cue in the GPT records equaled the true temperature in each round, so the manipulation again isolated the social channel rather than the environmental signal. Agents also answered the two post-task reconstruction questions analyzed here: recalled average temperature and recalled average own estimate. We parsed the GPT answers to these two questions to construct GPT analogues of the human post-event reconstruction variables. GPT agents did not receive any payoff; payoff variables were therefore not defined or analyzed. Exact instructions, post-task questions, model/run counts, and parsing rules are summarized in the SI.

\subsection{\texorpdfstring{Operationalizing the theory in the temperature task}{Operationalizing the theory in the temperature task}}

The Results present the minimal theory in compressed form. Here we give the operational definitions and derivations. In the human task, \(T_t\) was the displayed thermometer value, which equaled the true room temperature in the analysed sessions; \(x_i(t)\) was participant \(i\)'s un-noised submitted estimate; and \(S_i(t)\) was the social signal: the mean of the peer estimates on participant \(i\)'s screen. Signed prediction bias was computed as \(x_i(t)-T_t\).

Downward anchor displacement was defined as
\begin{equation}
A_i=T_1-x_i(1),
\label{eq:downward-anchor}
\end{equation}
so positive values indicate a first-round estimate below the evidence. Thus, negative signed bias in the figures corresponds to positive downward displacement in the theory.

The dynamic update models estimated Eq.~\ref{eq:dynamic-update} directly at the participant-round level. The outcome was \(x_i(t+1)-x_i(t)\). The coefficient on evidence pull, \(T_t-x_i(t)\), is the empirical estimate of the personal-evidence weight, \(\omega_i(t)\). The coefficient on peer pull, \(S_i(t)-x_i(t)\), is the empirical estimate of the social-information weight, \(\sigma_i(t)\). Peer dispersion, \(D_i(t)\), was the standard deviation of the peer estimates that formed the social signal, \(S_i(t)\). Confidence moderation models tested the social-information-weight structure in Eq.~\ref{eq:social-weight} by asking whether previous confidence changed the social-information weight, \(\sigma_i(t)\); \(\sigma_C<0\) means that higher confidence reduces social updating.

The same update equation also clarifies how evidence and social bias compete. Let \(b_i(t)=x_i(t)-T_t\) be current signed bias and let \(S_i^0(t)\) denote the faithful social signal without added channel noise. Equation~\ref{eq:dynamic-update} can be rewritten as
\begin{equation}
\Delta x_i(t)=-\omega_i(t)b_i(t)+\sigma_i(t)\,[S_i(t)-x_i(t)]+\epsilon_i(t),
\label{eq:bias-update}
\end{equation}
so positive personal-evidence weight pulls estimates back toward the evidence. In a faithful channel, the group-average social record carries the current group bias,
\begin{equation}
\bar S^0(t)-T_t=\bar x(t)-T_t=\bar b(t).
\label{eq:faithful-bias-record}
\end{equation}
With comprehension noise, \(S_i^C(t)=S_i^0(t)+\xi_i^C(t)\), where \(E[\xi_i^C(t)]=0\) and receiver-specific perturbations are approximately uncorrelated. With production noise, \(S_i^P(t)=S_i^0(t)+\xi_i^P(t)\), where perturbations are correlated across receivers because production-noise values are stored before display.

\subsubsection{\texorpdfstring{Technical note: variance and correlation of comprehension and production noise}{Technical note: variance and correlation of comprehension and production noise}}
\label{sec:channel-noise-technical-note}

For the covariance predictions, comprehension noise and production noise used the same integer perturbation support, \(-5\) to \(5\). The variance of a single perturbation was therefore \(\nu^2=10\). The difference between the two channels comes from how these perturbations enter the average social signal. Under comprehension noise, the group-average channel component contains \(N(N-1)\) independent receiver--production-noise perturbations, each with weight \(1/[N(N-1)]\). Summing squared weights gives
\begin{equation}
\mathrm{Var}(\bar{\eta}^{C})=N(N-1)\left[\frac{1}{N(N-1)}\right]^2\nu^2=\frac{\nu^2}{N(N-1)}.
\label{eq:comprehension-channel-variance}
\end{equation}
Under production noise, each production-noise perturbation is reused by the \(N-1\) receivers who observe that perturbed value. In the group-average social signal, each production-noise perturbation therefore has total weight \(1/N\). With \(N\) independent production-noise perturbations,
\begin{equation}
\mathrm{Var}(\bar{\eta}^{P})=N\left(\frac{1}{N}\right)^2\nu^2=\frac{\nu^2}{N}.
\label{eq:production-channel-variance}
\end{equation}
Dividing these two expressions gives
\begin{equation}
\frac{\mathrm{Var}(\bar{\eta}^{P})}{\mathrm{Var}(\bar{\eta}^{C})}=N-1,
\label{eq:variance-ratio}
\end{equation}
so production noise injects \(N-1\) times more variance into the group-average social signal than comprehension noise. With \(N=4\), Eq.~\ref{eq:variance-ratio} predicts \(10/12=0.833\) under comprehension noise and \(10/4=2.5\) under production noise.

Equation~\ref{eq:receiver-correlation} follows from the same production-noise covariance structure. Under production noise, the channel-noise component received by individual \(i\) is \((N-1)^{-1}\sum_{j\ne i}\eta_j^P\), and the component received by individual \(k\) is \((N-1)^{-1}\sum_{j\ne k}\eta_j^P\). These two sums share the \(N-2\) production-noise perturbations from group members other than \(i\) and \(k\). Their covariance is therefore \((N-2)\nu^2/(N-1)^2\), while each variance is \(\nu^2/(N-1)\). The correlation is
\begin{equation}
\mathrm{Corr}\!\left(S_i^{P},S_k^{P}\right)=
\frac{(N-2)\nu^2/(N-1)^2}{\nu^2/(N-1)}=\frac{N-2}{N-1}.
\label{eq:receiver-correlation}
\end{equation}
For four-person groups this gives \(2/3\). Under comprehension noise, receiver-specific perturbations are independent, so the corresponding receiver-to-receiver channel correlation is expected to be near zero.

\begin{table}[t]
\centering
\caption{Main variables used in the analysis.}
\label{tab:variables}
\begin{tabular}{p{0.22\textwidth}p{0.18\textwidth}p{0.52\textwidth}}
\toprule
Term & Symbol & Definition \\
\midrule
Prediction error & \(E_{\mathrm{truth}}\) & Mean absolute distance between estimate and actual temperature. \\
Signed prediction bias & \(x_i-T_t\) & Mean signed distance between estimate and actual temperature; negative values indicate underestimation. \\
Downward anchor displacement & \(A_i=T_1-x_i(1)\) & First-round evidence minus first-round estimate; positive values correspond to downward pre-social anchoring. \\
Evidence pull & \(T_t-x_i(t)\) & Evidence pull in the dynamic update model; its coefficient estimates the personal-evidence weight, \(\omega_i(t)\). \\
Peer pull & \(S_i(t)-x_i(t)\) & Peer pull in the dynamic update model; its coefficient estimates the social-information weight \(\sigma_i(t)\). \\
Social-signal inconsistency & \(E_{\mathrm{social}}\) & Mean absolute distance between the submitted estimate, \(x_i(t)\), and the social signal, \(S_i(t)\). \\
Social disagreement & \(D\) & Mean absolute distance between a participant and contemporaneous peers. \\
Confidence & -- & Mean reported confidence on the 0--10 scale. \\
Payoff & -- & \(\frac{5}{100}\sum_t \mathbf{1}(|\hat T_t-T_t|<10^\circ)/(|\hat T_t-T_t|+1)\). \\
Self-reconstruction error & \(\Delta_{\mathrm{self}}\) & Recalled average own estimate minus the data-derived average estimate. \\
Environment-reconstruction error & \(\Delta_{\mathrm{env}}\) & Recalled average temperature minus the actual average temperature. \\
Environment-self gap & \(\Delta_{\mathrm{env-self}}\) & Recalled average temperature minus recalled average own estimate. \\
Consensus alignment & \(M\) & \(\exp(-D/T_Q)\), high when participants agree; \(T_Q\) is the consensus-score tuning temperature. \\
Truth alignment & \(I\) & \(\exp(-E_{\mathrm{truth}}/T_Q)\), high when estimates are accurate. \\
Informative consensus & \(Q\) & \(M(2I-1)\), high only when agreement is accurate. \\
Harmful consensus & -- & \(D\leq5^\circ\) and \(E_{\mathrm{truth}}\geq5^\circ\). \\
\bottomrule
\end{tabular}
\end{table}

\section*{Acknowledgements}
We thank Iacoppo Hachen and Pavan Kaushik for fruitful discussions.

This work was supported by funding from the Deutsche Forschungsgemeinschaft (DFG, German Research Foundation) under Germany's Excellence Strategy -- EXC 2117-422037984, the Deutsche Forschungsgemeinschaft Gottfried Wilhelm Leibniz Prize 2022 584/22 (I.D.C.), the Max Planck Society, the European Union's Horizon 2020 Research and Innovation Programme under the Marie Sklodowska-Curie Grant agreement no. 860949, the Struktur- und Innovationsfonds fuer die Forschung of the State of Baden-Wuerttemberg, the PathFinder European Innovation Council Work Programme no. 101098722, the Office of Naval Research Grant N0001419-1-2556, and Small Project Grant (S23-32) from the Center for the Advanced Study of Collective Behavior.

\clearpage

\end{document}